\documentclass[11pt]{article}
\usepackage[T1]{fontenc}
\usepackage[utf8]{inputenc}
\usepackage{lmodern}
\usepackage[margin=1in]{geometry}
\usepackage{amsmath,amssymb,amsthm,mathtools}
\usepackage{microtype,enumitem,xurl,xcolor}
\usepackage[numbers,square,sort&compress]{natbib}
\usepackage{xcolor}
\usepackage[ruled,vlined,noend]{algorithm2e}
\newcommand{\ignore}[1]{}

\definecolor{algcommentcolor}{gray}{0.35}

\SetCommentSty{algcommentstyle}
\LinesNotNumbered
\usepackage[colorlinks=true,linkcolor=blue!40!black,citecolor=blue!40!black,urlcolor=blue!40!black]{hyperref}
\usepackage[nameinlink,noabbrev]{cleveref}
\crefname{algocf}{algorithm}{algorithms}
\Crefname{algocf}{Algorithm}{Algorithms}

\hypersetup{pdftitle={Local Computation and Estimation of Bipartite Maximum Matching},
pdfauthor={Edith Cohen and Mohammad Roghani}}
\numberwithin{equation}{section}
\newtheorem{theorem}{Theorem}[section]
\newtheorem{lemma}[theorem]{Lemma}
\newtheorem{proposition}[theorem]{Proposition}
\newtheorem{corollary}[theorem]{Corollary}
\theoremstyle{definition}
\newtheorem{definition}[theorem]{Definition}
\theoremstyle{remark}
\newtheorem{remark}[theorem]{Remark}
\newcommand{\E}{\mathbb E}

\newcommand{\eps}{\varepsilon}
\newcommand{\Prob}{\mathbb{P}}
\newcommand{\R}{\mathbb{R}}
\newcommand{\cN}{\mathcal{N}}
\newcommand{\dmy}{\bot}
\newcommand{\inner}[2]{\langle #1,#2\rangle}
\DeclareMathOperator{\osc}{osc}
\DeclareMathOperator{\Resp}{Resp}
\DeclareMathOperator{\atanh}{atanh}
\DeclareMathOperator{\Out}{Out}
\DeclareMathOperator{\In}{In}
\DeclareMathOperator{\MD}{MD}
\newcommand{\coin}[1]{\mathsf{#1}}

\title{Optimal LCA and Sublinear Estimation \\for Bipartite Matching in \texorpdfstring{$\Delta^{O(1/\varepsilon)}$}{Delta\^{}\{O(1/epsilon)\}} Queries}
\author{Edith Cohen\\
Google Research and Tel Aviv University\\
\texttt{edith@cohenwang.com}
\and
Mohammad Roghani\\
Google Research\\
\texttt{roghani@google.com}}
\date{}

\begin{document}

\maketitle

\begin{abstract}
A local computation algorithm (LCA) for matching answers
``who is the mate of $v$?'' by inspecting only a small
part of the graph, with all answers consistent with one
matching. For $n$-vertex graphs of maximum degree
$\Delta\ge2$, Levi et al. [Algorithmica'17] give a
$(1-\varepsilon)$-approximate matching LCA using
$(\Delta/\varepsilon)^{O(1/\varepsilon^2)}
\operatorname{polylog}(n)$ queries per invocation,
while Yoshida et al. [SIAM J. Comput.'12] estimate
the maximum matching size within additive error
$\varepsilon n$ using $\Delta^{O(1/\varepsilon^2)}$
queries. Behnezhad, Roghani, and Rubinstein [FOCS'23]
proved a $\Delta^{\Omega(1/\varepsilon)}$ lower bound
for both tasks, even on bipartite graphs and even when
the LCA's matching guarantee is relaxed to additive
$\varepsilon n$ error. This leaves a gap between linear
and quadratic dependence on $1/\varepsilon$ in the exponent.

We close this gap on bipartite graphs. We give an LCA that answers each mate query with $\Delta^{O(1/\varepsilon)}$ queries for a $(1-\varepsilon)$-approximate matching, and an estimator for the maximum matching size within $\varepsilon n$ using $\Delta^{O(1/\varepsilon)}$ queries. On general graphs, the same algorithm gives a half-integral fractional matching and estimates the maximum fractional matching size.

Our approach runs $O(\varepsilon^{-1}\log\Delta)$ rounds
of a deterministic local optimization method in which
vertices propose to neighbors and update prices according
to incoming loads. We design its updates to admit exact
recursive sampling with constant expected branching,
obtaining the improved randomized bound without exploring
full dependency neighborhoods. To our knowledge, this is
the first LCA construction based on exact recursive sampling
of the iterates of a deterministic continuous optimization
method. We expect this approach to have applications
beyond matching.
\end{abstract}

\section{Introduction}

A \emph{matching} in a graph $G=(V,E)$ is a set of edges
no two of which share an endpoint, and $\nu(G)$ denotes
its maximum size. Maximum matching is a central problem
in combinatorial optimization, with applications ranging
from assignment problems to sparse matrix computation
\citep{Kuh55,DKU11}. Its algorithmic study spans more than
six decades, with Edmonds's polynomial-time algorithm for
general graphs as a foundational milestone \citep{Edm65}.

We study graphs with $n$ vertices and maximum degree at
most $\Delta$, where $n$ may be much larger than $\Delta$.
For such graphs, even reading the entire input may be
too costly. Our goal is to answer local queries about an
almost maximum matching, or to estimate its size, while
inspecting only a small part of the graph.
The first task is the subject of \emph{local computation
algorithms} (LCAs), introduced by
\citeauthor*{RTVX11}~\cite{RTVX11} and
\citeauthor*{ARVX12}~\cite{ARVX12}.
The second is sublinear-time estimation, studied since
the work of \citeauthor*{PR07}~\cite{PR07} and
\citeauthor*{NO08}~\cite{NO08}.
Matching has received extensive attention in both
settings~\cite{PR07,NO08,YYI12,MV13,LRY17, BRR23, brr23a, brr24, BKS23a, BKS23b, ABRR25, shah26}, and we review the known bounds below.
We also consider the distributed LOCAL and CONGEST models,
where vertices communicate with their neighbors in
synchronous rounds and complexity is measured by the
number of communication rounds.

We study both matchings and their classical fractional
relaxation. A \emph{fractional matching} assigns nonnegative
weights $f_e$ to edges such that $\sum_{e\ni v}f_e\le1$
at every vertex. Its value is $\sum_e f_e$, and $\nu_f(G)$
denotes the maximum possible value. It is \emph{half-integral}
if every weight belongs to $\{0,\tfrac12,1\}$.
On bipartite graphs, $\nu_f(G)=\nu(G)$.

\paragraph{The local computation model.}
The vertices are labeled by $[n]$, and $n$ and a degree
bound $\Delta\ge2$ are known.
A degree query at $v$ returns $\deg(v)$.
A neighbor query at $(v,i)$ returns the $i$-th neighbor
of $v$, with neighbors listed in a fixed but arbitrary
order.
An LCA~\cite{ARVX12,RTVX11} provides query access to a
solution without constructing the entire output.
For matching, a \emph{mate query} at a vertex returns
its partner or reports that it is unmatched.
For fractional matching, an \emph{edge query} returns
the weight assigned to an edge.

Our LCAs do not use preprocessing and do not retain state
between invocations. A randomized LCA has access to
a random tape shared by all invocations, allowing
repeated evaluations to reproduce the same random
choices. All answers must be globally consistent:
they must agree with one feasible solution, in every
query order, including adaptively chosen queries.

We measure an LCA's query complexity by the worst-case
number of graph queries used to answer one solution query.
A \emph{size estimator} returns a numerical estimate of
$\nu(G)$ or $\nu_f(G)$, and its query complexity is the
total number of graph queries it makes.

Computation is free when measuring graph-query complexity.
We separately bound computational work in the finite-arithmetic model of \Cref{sec:model}.

\paragraph{The distributed LOCAL and CONGEST models.}
In both models \citep{Pel00}, each vertex is a processor that
communicates with its neighbors in synchronous rounds.
The LOCAL
model measures information radius and allows messages of unrestricted size; the CONGEST additionally restricts communication bandwidth to
$O(\log n)$ bits per edge in each direction
per round.
After $r$ rounds, a vertex's output can depend only on
information within its radius-$r$ neighborhood.
Thus distributed round complexity measures how far information must
travel, whereas LCA query complexity measures how much
of the input is inspected.
An LCA can simulate an $r$-round distributed algorithm by
exploring the radius-$r$ neighborhood of the queried
vertex, using $(2\Delta)^{O(r)}$ graph queries
\citep{PR07}. In addition to query and round complexity, we separately
bound the computational work per local query and the
message size per edge per round in our distributed
implementation.

\paragraph{Approximation guarantees.}
The optimal matching can depend on information arbitrarily
far from the queried vertex, even on paths. We therefore
focus on approximation when seeking bounds independent
of $n$. 

For the matching itself, we seek a multiplicative guarantee:
a $(1-\eps)$-approximate matching has size at least
$(1-\eps)\nu(G)$, and analogously at least $(1-\eps)\nu_f(G)$ for fractional matching.
To describe earlier results, we also use additive
approximation notation: a matching $M$ is
$(1,\eps n)$-approximate if $|M|\ge\nu(G)-\eps n$,
and more generally $(\alpha,\eps n)$-approximate if
$|M|\ge\alpha\nu(G)-\eps n$. Note that since
$\nu(G),\nu_f(G)\le n/2$, a multiplicative $(1-\eps)$-approximate guarantee implies an
additive $(1,\eps n/2)$ guarantee and hence is stronger.

For size estimation, a multiplicative guarantee with a fixed $\eps$
requires distinguishing the empty graph from a graph
with a single edge, which requires $\Omega(n)$ queries. For query bounds independent of $n$, we therefore seek additive accuracy.

\paragraph{Known algorithms.}
Sublinear algorithms for estimating matching size in
bounded-degree graphs go back to
\citeauthor*{PR07}~\cite{PR07}, who showed how to simulate
a distributed algorithm at a single vertex by exploring
its neighborhood. For matching, this gives a
$(1/2,\eps n)$-estimate with
$\Delta^{O(\log(\Delta/\eps))}$ queries.
\citeauthor*{NO08}~\cite{NO08} introduced recursive local
simulation of randomized greedy algorithms, and
\citeauthor*{YYI12}~\cite{YYI12} improved this approach
to obtain a $(1,\eps n)$-estimate with
$\Delta^{O(1/\eps^2)}$ queries.
Applied to the standard bipartite double cover
\citep[Lemma~7]{AKO18}, this estimator also estimates
$\nu_f(G)$ within $\pm\eps n$ with the same query bound.

For LCAs, \citeauthor*{MV13}~\cite{MV13} gave an
approximation scheme with polylogarithmic query complexity
for constant $\Delta$ and $\eps$.
\citeauthor*{LRY17}~\cite{LRY17} adapted the method
of~\cite{YYI12} to obtain a multiplicative
$(1-\eps)$-approximation using
$(\Delta/\eps)^{O(1/\eps^2)}\operatorname{polylog}(n)$
queries per invocation. This remains the best known
bound of this form for general-graph matching LCAs.
These approximation schemes use the augmenting-path
framework of Hopcroft and Karp: in each of
$\Theta(1/\eps)$ phases, they augment along a maximal
collection of vertex-disjoint paths of a prescribed
length. Finding this collection amounts to computing
a maximal independent set in the conflict graph of
candidate paths. In the simulations of~\cite{YYI12,LRY17},
each phase recursively invokes earlier phases, and
the costs compound to give the quadratic dependence
on $1/\eps$ in the degree exponent.

A parallel line of work seeks sublinear algorithms
without assuming a small maximum degree.
For fixed $\eps$, 
\citeauthor*{behnezhad2022time}~\cite{behnezhad2022time}
gave a $(1/2,\eps n)$-estimate of matching size in
$\tilde O(\bar d+1)$ time with adjacency-list queries,
where $\bar d$ is the average degree, and in
$\tilde O(n)$ time with adjacency-matrix queries.
A multiplicative $(1/2-\eps)$-estimate is also possible
in $\tilde O(n)$ time with adjacency-list queries.
All three bounds are optimal up to polylogarithmic
factors for fixed $\eps$.
Subsequent work improved the approximation ratio
in $n^{2-\Omega(1)}$ time
\cite{behnezhad2023beating,brr23a,BKS23a}.
Finally, \citeauthor*{BKS23b}~\cite{BKS23b}
obtained a $(1,\eps n)$-estimate in
$n^{2-\Omega_\eps(1)}$ time using adjacency-matrix queries.

\paragraph{Known lower bounds.}
For over fifteen years, the only degree-dependent lower bound
for matching LCAs was the $\Omega(\Delta)$ bound of
\citeauthor*{PR07}~\cite{PR07}, based on hiding useful
matching edges among many irrelevant neighbors.
This hiding argument alone cannot give a superlinear
dependence on $\Delta$.
\citeauthor*{brr23a}~\cite{brr23a} first broke this barrier,
showing that $(2/3+\Omega(1),\eps n)$-estimation requires
$n^{1.2-o(1)}$ queries for sufficiently small constant
$\eps$. Their construction, with $\Delta=\Theta(n)$,
also yields a $\Delta^{1.2-o(1)}$ lower bound per LCA query.

Open Problem~39 of sublinear.info~\cite{OP39} asked
whether $\operatorname{poly}(\Delta/\eps)$ queries
suffice to estimate matching size within additive error
$\eps n$. \citeauthor*{BRR23}~\cite{BRR23} answered this negatively,
proving a $\Delta^{\Omega(1/\eps)}$ lower bound for both
estimation and local computation, even on bipartite
graphs, whenever $\eps\le0.01$ and
$\log^4 n\le\Delta\le n^\eps$.

For dense graphs, \citeauthor*{brr24}\cite{brr24} subsequently proved
that, for every fixed $\delta>0$, there is a constant
$\eps=\eps(\delta)>0$ such that additive $\eps n$
estimation requires $\Omega(n^{2-\delta})$
adjacency-list queries. This complements the
$n^{2-\Omega_\eps(1)}$ upper bound of~\cite{BKS23b}.
In the bounded-degree regime, the lower bound
of~\cite{BRR23} left a gap between linear and quadratic
dependence on $1/\eps$ in the degree exponent.


\subsection{Our results} 

Our randomized algorithms close this gap for bipartite
matching and its size estimation. On general graphs, they
give the same bounds for half-integral fractional matching
and estimation of $\nu_f(G)$.

\begin{theorem}[Randomized local computation and estimation,
informal]\label{thm:randomized}
Let $G$ have $n\ge2$ vertices and maximum degree at most
$\Delta\ge2$, and let $0<\eps<1/2$.
\begin{enumerate}[label=(\alph*)]
\item
There is a single randomized LCA with the following
guarantees. On bipartite graphs, it supports mate queries
to a matching $M$ satisfying
\[
\Prob\left\{|M|\ge(1-\eps)\nu(G)\right\}\ge0.9.
\]
On arbitrary graphs, the same LCA supports edge queries
to a half-integral fractional matching $f$ of value
at least $(1-\eps)\nu_f(G)$ with probability at least $0.9$.

Each mate or edge query uses $(2\Delta)^{O(1/\eps)}$
graph queries. No bipartition or prior knowledge of
whether $G$ is bipartite is required.

\item
For every $0<\zeta<1/2$, a randomized size estimator
returns $\widehat\nu$ satisfying
$|\widehat\nu-\nu_f(G)|\le\eps n$
with probability at least $1-\zeta$, using
\[
(2\Delta)^{O(1/\eps)}\log(2/\zeta)
\]
graph queries.
On bipartite graphs, this estimates $\nu(G)$.
\end{enumerate}
The same bounds hold separately for computational work
in the model of \Cref{sec:model}.
All query and work bounds are worst case.
The LCA's feasibility and consistency hold for every
realization of the shared randomness and in every query
order, including adaptively chosen queries.
Only the approximation guarantees are probabilistic.
\end{theorem}

For constant success probability, our upper bounds match
the lower bound of \citet{BRR23}, in its applicable regime,
up to a constant in the exponent. Thus, in that regime,
the query complexity of bipartite matching and its size
estimation is $\Delta^{\Theta(1/\eps)}$.

For general graphs, $\nu(G)\le\nu_f(G)\le\frac32\nu(G)$.
Consequently, scaling the estimate in part~(b) by $2/3$
gives a $(2/3,\eps n)$-estimate of $\nu(G)$:
with probability at least $1-\zeta$,
\[
\frac23\nu(G)-\eps n
\le \frac23\widehat\nu
\le \nu(G)+\eps n.
\]
Whether integral $(1,\eps n)$-approximate matchings in
general graphs admit an LCA using
$(2\Delta)^{O(1/\eps)}$ graph queries per mate query
remains the main open question left by our work.
The lower bound of \citet{BRR23} applies, while the
general-graph algorithms of \citet{YYI12,LRY17} have
quadratic dependence on $1/\eps$ in the degree exponent.

Our construction also gives a deterministic LCA for
fractional matching, with a larger query bound.

\begin{theorem}[Deterministic fractional oracle,
informal]\label{thm:deterministic}
Let $G,\Delta,\eps$ be as in \Cref{thm:randomized}.
There is a deterministic LCA whose answers are consistent
with one fractional matching $f$ satisfying
\[
\sum_e f_e\ge(1-\eps)\nu_f(G).
\]
Each edge query uses at most
\[
(2\Delta)^{O(\eps^{-1}\log(2\Delta))}
\]
graph queries, all within distance
$O(\eps^{-1}\log(2\Delta))$ of the queried edge.

For every $0<\zeta<1/2$, uniform random vertex sampling
combined with this oracle gives an estimate of $\nu_f(G)$
within $\pm\eps n$, with probability at least $1-\zeta$,
using at most
\[
(2\Delta)^{O(\eps^{-1}\log(2\Delta))}\log(2/\zeta)
\]
graph queries in total. Only the vertex sampling in this
estimator is randomized.

The same bounds hold separately for computational work
in the model of \Cref{sec:model}.
All query and work bounds are worst case.
\end{theorem}

The locality and structure of the deterministic updates give the
following distributed consequence.

\begin{corollary}[Deterministic distributed algorithm,
informal]\label{cor:local-intro}
Under the same assumptions, a deterministic distributed
algorithm computes a fractional matching of value at least
$(1-\eps)\nu_f(G)$ in
\[
O\!\left(\frac{\log(2\Delta)}{\eps}\right)
\]
rounds. Each edge carries $O(\log(2\Delta/\eps))$ bits
in each direction per round.
The total computational work at a vertex $v$ is
\[
(\deg(v)+1)\operatorname{poly}\!\left(
\frac{\log(2\Delta)}{\eps}\right).
\]
All bounds are worst case.
This gives the stated round bound in LOCAL for every
$0<\eps<1/2$, and in CONGEST when $\eps\ge n^{-O(1)}$.
\end{corollary}

Fractional matching is a special case of fractional
packing, which has been studied in distributed computing
\citep{KMW16}. The deterministic algorithm of
\citet{LOW24} gives a $(1-\eps)$-approximate fractional
matching in $O(\eps^{-2}\log(2\Delta))$ rounds.
Simulating it by neighborhood exploration \citep{PR07}
gives a deterministic LCA using
$(2\Delta)^{O(\eps^{-2}\log(2\Delta))}$
graph queries per edge query.
Compared with these bounds, our construction improves
the dependence on $1/\eps$ from quadratic to linear,
both in the LOCAL round complexity and in the exponent
of the LCA query complexity.

\subsection{Technical overview and roadmap}\label{sec:approach}

We first formulate fractional matching as a saddle-point
problem and derive a deterministic optimization method with
local updates (\Crefrange{sec:game}{sec:opt}).
After $T=O(\eps^{-1}\log(2\Delta))$ rounds, the method
defines a nearly maximum fractional matching whose edge
weights depend only on radius-$O(T)$ neighborhoods.
Evaluating the updates in these neighborhoods gives a
deterministic LCA using
$(2\Delta)^{O(T)}=(2\Delta)^{O(\eps^{-1}\log(2\Delta))}$
queries and operations per edge query, and a deterministic
distributed algorithm using $O(T)$ communication rounds.

The key to improving the LCA bound is to sample from the
distributions used by the optimizer without computing them
explicitly. In \Cref{sec:sampling}, we construct a recursive
sampler that follows the optimization round by round.
Producing a sample for a new round uses an expected constant
number of calls to the preceding round's sampling procedures.
This replaces the recursive exploration of all neighbors
by constant expected branching, giving
$2^{O(T)}=(2\Delta)^{O(1/\eps)}$ expected queries and
operations per sample.

We then use these samples to construct a fractional matching
that can be queried consistently with
$(2\Delta)^{O(1/\eps)}$ worst-case work per query
(\Cref{sec:prop}).
Rounding gives an integral matching on bipartite graphs
and a half-integral fractional matching on general graphs
(\Cref{sec:lca}).
The same sampler also gives our size estimator, without
the rounding step (\Cref{sec:est}). The finite-precision implementation and its message-size
bound are established in \Cref{sec:finite-precision}.
We next describe each ingredient and the challenges it addresses.

\subsubsection{A saddle-point formulation of fractional matching}

In \Cref{sec:game}, we express fractional matching as a
min--max problem over distributions and prices at the
vertices. Every vertex acts both as a proposer, sending
mass to its neighbors, and as a receiver, accepting mass
from its neighbors. Its proposal distribution describes
the mass it sends, while its price is associated with
its role as a receiver.

Each vertex $u$ distributes one unit of mass among its
neighbors and a private dummy, which represents unused
mass. We call this probability vector $q_u$ its
\emph{proposal distribution}. The quantities $q_{uv}$
and $q_{vu}$ represent mass sent in opposite directions
and need not agree. Each vertex sends at most one unit
to its neighbors, but may receive more than one unit.
Write
\[
\ell_v(q)=\sum_{u\in N(v)}q_{uv},\qquad
P(q)=\sum_v\min\{1,\ell_v(q)\}
\]
for the incoming load at $v$ and the total accepted mass
when every vertex can accept at most one unit.

To obtain a feasible fractional matching, divide all
incoming assignments at each vertex $v$ with
$\ell_v(q)>1$ by $\ell_v(q)$. Every vertex now sends
and receives at most one unit. Assign each undirected
edge the average of its two accepted directed assignments.
The incident weights at each vertex sum to at most one,
and the total edge weight is $P(q)/2$. Conversely, any
fractional matching defines proposal distributions by
sending its edge weights in both directions and putting
unused mass on dummies. The resulting accepted mass is
twice the matching's value. Thus
$\max_q P(q)=2\nu_f(G)$.

The cap makes $P(q)$ nonsmooth. We introduce a price
$x_v\in[-1,1]$ at each vertex $v$, associated with its
incoming load, and use the objective
\[
F(x,q)=\sum_v\bigl[x_v+\ell_v(q)(1-x_v)\bigr],
\qquad
\min_x F(x,q)=2P(q)-n.
\]
The second identity follows by minimizing separately
at each vertex: an overloaded vertex chooses price
$+1$, and an underloaded vertex chooses price $-1$.
For fixed prices, proposal mass prefers lower-priced
neighbors. Maximizing accepted mass is therefore
equivalent to maximizing over proposal distributions
and minimizing over prices. This is a
\emph{saddle-point problem}, with an objective affine
in either set of variables when the other is fixed.

The multi-pass semi-streaming algorithm of
\citet{AJJST22} also uses a saddle-point formulation
of matching. Their formulation uses a single simplex
over all edges, scaled by an estimate of the optimum,
so normalization involves the whole graph.
Our formulation uses a product of simplices, one
for each vertex. Each proposal distribution is
normalized separately at its source, which allows
the local updates described next. The formulation
also does not require an estimate of $\nu_f(G)$.

\subsubsection{Optimization via local updates}

\Crefrange{sec:reg}{sec:opt} develop a deterministic
optimization method for nearly maximizing accepted mass.
Observe that simply alternating between best proposal distributions
and best prices can oscillate the state: proposal mass moves to
cheap receivers, making them overloaded and expensive,
and then moves away again. 
We control these changes by penalizing
movement away from the current state using a
\emph{Bregman divergence} derived from our regularizer.
The regularizer combines negative entropy of the proposal
distributions with an entropy-based price term that depends
on the incoming loads.

We initialize each nonisolated vertex with price
$-1+1/\Delta$ and assign probability $1/(2\Delta^2)$
to each of its neighbors, putting the remaining
probability on its dummy.

Starting from this state, we perform $T$ rounds of an
extragradient method \citep{Nem04}. Each round first
computes a tentative state using the gradient at the
current state. It then evaluates the gradient at the
tentative state and uses this gradient to form the
next state, again starting from the original state.
Both stages use penalties measured relative to the
state at the start of the round.

The proposal updates have a multiplicative-weights form,
with normalization performed separately at each vertex.
Prices are updated by explicit formulas using the old
prices and the relevant loads. The final output
$\widehat q$ is the average of the tentative proposal
distributions over all $T$ rounds.

Our regularizer satisfies \emph{area convexity}
\citep{She17,JT23}, an inequality controlling the
interaction between changes in prices and changes
in proposal distributions. With the initialization
above, the analysis of the two-stage method gives
\[
2\nu_f(G)-P(\widehat q)
\le O\!\left(\frac{\nu_f(G)\log(2\Delta)}{T}\right).
\]
Since $\widehat q$ defines a fractional matching of
value $P(\widehat q)/2$, taking
$T=O(\eps^{-1}\log(2\Delta))$ gives a
$(1-\eps)$-approximate fractional matching.
Note that the optimization, including the choice of $T$, does not require an
estimate of $\nu_f(G)$.

The state at each vertex in a new round depends only
on the preceding round's states within a constant
number of hops. Since the initialization also uses
only local information, each final edge weight is
determined by a radius-$O(T)$ neighborhood.

A deterministic LCA can explore this neighborhood
and evaluate the updates, using
\[
(2\Delta)^{O(T)}
=(2\Delta)^{O(\eps^{-1}\log(2\Delta))}
\]
graph queries and arithmetic operations per edge query.
This gives the deterministic LCA bound in
\Cref{thm:deterministic}.
Each optimization round uses a constant number of exchanges
between neighboring vertices. Executing these updates gives
an $O(T)=O(\eps^{-1}\log(2\Delta))$-round distributed
algorithm. The finite implementation in
\Cref{sec:finite-optimizer} bounds the message sizes and
local computational work, giving the distributed guarantees
in \Cref{cor:local-intro}.

The regularizer is also designed to support the 
sampling method described next.

\subsubsection{Sampling: following the optimization without evaluating it}

The deterministic LCA computes $\widehat q_u$ explicitly
by exploring an entire radius-$O(T)$ neighborhood of $u$,
containing up to $\Delta^{O(T)}$ vertices.
In \Cref{sec:sampling}, we show how to draw an \emph{exact}
sample from $\widehat q_u$ without computing its
probabilities, using only $2^{O(T)}$ expected graph
queries and arithmetic operations.

Our multiplicative-weights updates form a new proposal
distribution $h_u$ by reweighting an existing distribution
$q_u$:
\[
h_{ua}=\frac{q_{ua}A_{ua}}{Z_u},\qquad
Z_u=\sum_b q_{ub}A_{ub}.
\]
By design, the reweighting factors in our update formulas
have fixed positive lower and upper bounds. Multiplying
them all by a common factor leaves $h_u$ unchanged.
We can therefore choose a fixed rescaling that puts
every $A_{ua}$ between two constants in $(0,1)$,
so these weights can serve as acceptance probabilities.

Suppose we have a sampler for $q_u$ and, for any sampled
target $a$, an \emph{acceptance coin} that returns one
with probability $A_{ua}$. Draw $a\sim q_u$ and flip
this coin. On heads, return $a$; on tails, repeat with
a fresh draw and coin flip. A trial accepts and returns
$a$ with probability $q_{ua}A_{ua}$, so the returned
sample has law $h_u$. The acceptance probability $Z_u$
is bounded below by a constant, giving a constant
expected number of trials. Thus an existing sampler
and acceptance coins suffice to construct the new
sampler, without evaluating either distribution or
its normalizer.

The remaining task is to construct the acceptance
coins. Their biases depend on prices and loads, and
evaluating these quantities numerically would bring
back the recursive exploration of all neighbors.
We instead maintain coin procedures for the numerical
quantities needed by the updates. For example, a price
$x_v\in[-1,1]$ can be represented by a coin whose
probability of heads is $(2+x_v)/4$. The update formulas
can then be expressed as functions of these coin biases.

These constructions use \emph{Bernoulli factories}.
Given independent flips of coins with unknown biases,
a factory produces a new coin whose bias is a prescribed
function of those biases
\citep{KO94,NP05,HNP11,Huber16,Mendo19}.
We use factories to implement the scalar update formulas
on coin representations. A new acceptance or price coin
is constructed from available coins and samplers.
Each input flip uses fresh randomness and may itself
invoke earlier procedures.

The price updates also use incoming loads.
At a nonisolated vertex $v$, the incoming proposal masses
$q_{uv}$ sum to the load $\ell_v(q)$, which may not be equal to one.
Normalizing these masses defines the \emph{incoming
distribution}, assigning probability $q_{uv}/\ell_v(q)$
to proposer $u$. We maintain a sampler for this
distribution as well.

When $q$ changes to $h$, the incoming distribution
is reweighted by $h_{uv}/q_{uv}$. Factories construct
coins for scaled versions of these ratios from the
available samplers and acceptance coins. The same
rejection-sampling construction then gives the new
incoming sampler. An incoming rejection trial also
supplies a coin whose bias is a fixed constant times
the ratio of the new load to the old load. Thus the
construction gives both samplers and load-ratio coins.
Factories combine these load coins with coins for the
old state to construct the updated price coins.

Constant expected branching requires a constant
number of rejection trials, which is given by the bounded reweighting factors, and a constant expected
number of input flips per factory, which follow from our regularizer design and coin representations together with a constant
expected number of input flips and additional arithmetic
operations.  These bounds
hold uniformly over vertices and rounds.

At initialization, all the required samplers and coins
are directly available. For each new round, one sample
or coin flip uses an expected constant number of calls
to the preceding round's procedures. Expanding these
calls gives expected total work
\[
2^{O(T)}=(2\Delta)^{O(1/\eps)},
\]
including graph queries and arithmetic operations.
The calls stay within a radius-$O(T)$ neighborhood,
exploring it only as needed to generate the draw.
To sample from the optimizer's final average
$\widehat q_u$, we choose a round uniformly from
the $T$ rounds and sample its tentative distribution
at $u$.

\subsubsection{From samples to a consistent fractional matching}

\Cref{sec:prop} uses $K=\Theta(\eps^{-2})$ samples per
vertex to construct a random fractional matching whose
edge weights can be queried consistently with
$(2\Delta)^{O(1/\eps)}$ worst-case graph queries and
arithmetic operations.

The first issue is that the sampler's work is bounded
only in expectation, whereas the final oracle requires
a worst-case bound. We address this by imposing a work
budget on each sampler invocation, with aborted draws
returning the dummy. By Markov's inequality, a budget
of $\Theta(\Delta/\eps)$ times the bound on expected
work makes the abort probability $O(\eps/\Delta)$.
Let $\tilde q$ denote the resulting proposal
distributions. For every real edge $uv$,
$\tilde q_{uv}\le\widehat q_{uv}$, with the removed
probability mass transferred to the dummy.
If $n_0$ is the number of nonisolated vertices, the
work budget reduces accepted mass by at most
$O(\eps n_0/\Delta)$. Since
$n_0\le2\Delta\nu_f(G)$ (\Cref{lem:active}),
this loss is $O(\eps\nu_f(G))$.

We next construct a random fractional matching on $G$
from samples of $\tilde q$.
Let each vertex $u$ send $K$ independent proposals
drawn from $\tilde q_u$, ignoring dummy outcomes.
Each receiver keeps the $K$ proposals with smallest
IDs, or all its proposals if $K$ or fewer arrive.
Regard each kept proposal as a separate edge copy in
a multigraph and give it weight $1/(2K)$.
Every vertex has at most $K$ kept outgoing copies
and $K$ kept incoming copies, so these weights always
form a fractional matching. Summing the weights of
parallel copies gives a random fractional matching
on $G$. 

We bound the expected loss from discarding proposals
at each receiver relative to its accepted mass under
$\tilde q$. If receiver $v$ has incoming load
$\ell=\ell_v(\tilde q)$ and receives $N$ proposals,
then $\E N=K\ell$ and
\[
\frac{\E\min\{K,N\}}K
\ge\bigl(1-O(K^{-1/2})\bigr)\min\{1,\ell\}.
\]
Summing over receivers, the expected fractional value
is therefore at least
$(1-O(K^{-1/2}))P(\tilde q)/2$.
The earlier work-budget bound gives
$P(\tilde q)\ge P(\widehat q)-O(\eps\nu_f(G))$,
and the deterministic optimizer guarantees
$P(\widehat q)/2\ge(1-O(\eps))\nu_f(G)$.
With $K=\Theta(\eps^{-2})$, the three losses---from
sampling and the receiver cap, aborted sampler runs,
and the optimizer's approximation error---sum to
$O(\eps\nu_f(G))$. Thus the resulting random fractional
matching has expected value at least
$(1-O(\eps))\nu_f(G)$.

For consistency in an LCA implementation, give each
proposal ID its own independent random seed.
Whenever a query needs that proposal, it reruns the
same capped invocation with the same seed.
The receiver decisions are determined by the proposal
IDs, so all queries access one multigraph of kept
proposals, regardless of their order.

Finding the proposals received by a vertex requires
examining its neighbors and their $K$ proposals.
Listing all kept copies incident to the vertex also
checks which of its outgoing proposals were kept by
their receivers. This takes $O(\Delta K^2)$ capped
sampler invocations. The neighborhood exploration
occurs outside the recursive sampler and adds only
a polynomial overhead in $\Delta$ and $1/\eps$.
Together with the work budget, this gives the stated
$(2\Delta)^{O(1/\eps)}$ worst-case work per local query.

For the rounding in \Cref{sec:lca}, we choose $K$ to
be a power of two, preserving $K=\Theta(\eps^{-2})$.

\subsubsection{Rounding: integral on bipartite graphs,
half-integral in general}

The kept proposals define a multigraph $H$, with one
edge copy per kept proposal. Its maximum degree is
at most $D=2K$, and each edge copy has weight $1/D$.
Recall that $K$ was chosen as a power of two, so $D$
is also a power of two. In \Cref{sec:lca}, we adapt
the degree-halving methods of \citet{Fis17,AKO18}
to round this representation with bounded local work.
Each step halves the degree bound and doubles the
weight of every remaining edge copy.

First consider a bipartite multigraph. At each vertex,
pair incident copies in ID order, leaving one unpaired
if the degree is odd. These pairings link the copies
into paths and even cycles. Keeping the larger of
the two alternating classes of copies in each
component retains at least half the copies while
halving the degree bound. After doubling the remaining
weights, each pair contributes the same total weight
at its shared endpoint. The halved degree bound
ensures that every vertex's total incident weight
remains at most one, including vertices with an
unpaired copy. Thus this step preserves feasibility
without decreasing the total value.

The paths and cycles may be long, so deciding which
class to keep cannot require exploring an entire
component. This is addressed by deleting sparsely marked copies to break
these components into short pieces. A further local
rule deletes copies whose sufficiently large
neighborhood along the path or cycle contains no mark.
This ensures bounded exploration for every choice
of marks. We choose the parameters so that the
accumulated expected loss from deletions over the
$\log_2 D=O(\log(2/\eps))$ halving steps is at most
an $O(\eps)$ fraction of the starting value.
The remaining copies then form a matching.
The extra factor in worst-case work per query is
$\exp(O(\log^2(2/\eps)))$, which is absorbed into
our LCA bound.

To handle all graphs with one construction, we apply
this rounding to the \emph{bipartite double cover}
of $H$. It has a left and a right copy of each
vertex and replaces each edge copy $uv$ of $H$
by two lifted edges, $u_Lv_R$ and $v_Lu_R$.
Each matched lifted edge contributes weight $1/2$
to its original edge, using the standard double-cover
projection \citep[Lemma~7]{AKO18}.
Each original vertex has two lifts, each incident
to at most one matched edge. Thus, after summing
contributions from parallel copies, the projected
weights are feasible and belong to $\{0,1/2,1\}$.
The double cover has twice the initial fractional
value, and projection divides the rounded value
by two, so the relative-loss bound is unchanged.

The two lifts of each edge copy of $H$ share the
same random mark at every level. Pairings and ties
are resolved by ordering lifted copies first by
their original IDs, with a side index used only
to distinguish the two lifts of the same copy.
For bipartite input, the double cover consists of
two isomorphic copies of each connected component
of $H$. Each original edge ID appears only once
in either component copy, so the secondary side
indices do not affect the ordering there.
The shared marks and corresponding orders therefore
give corresponding rounding decisions: each edge
copy of $H$ has either both lifts in the final
matching or neither. Its projected weight is
therefore integral. The algorithm gives a matching
on bipartite graphs without receiving a bipartition
or testing bipartiteness. On general graphs it gives
a half-integral fractional matching.

Note that feasibility and consistency hold for every choice
of the budget-capped proposal samples and rounding marks.
Only the value guarantee is probabilistic.
By assigning a sufficiently small constant fraction
of the error allowance to each stage, we make the
expected deficit from the optimum at most
$\eps\nu_f(G)/10$. Markov's inequality then gives
a single consistent global output of value at least
$(1-\eps)\nu_f(G)$ with probability at least $0.9$.
On bipartite graphs, $\nu_f(G)=\nu(G)$ and the output
is integral, giving the claimed matching guarantee.

\subsubsection{Size estimation: sample the accepted mass.}
The size of the matching can be estimated by a black box use of our LCA by uniformly 
sampling random vertices and querying their total incident matching weight.  Multiplying the average by $n/2$
estimates the value of the LCA's output, which approximates
$\nu_f(G)$ whenever its value guarantee holds.
In \Cref{sec:est} we present an algorithm  with 
a simpler construction, which requires only 
the budget-capped sampler, with fresh independent
randomness in each trial. In particular, it does not require rounding
or consistency across queries.
The direct approach estimates the expected mass accepted
at a uniformly random receiver over the random construction
of $H$, using only capped proposal samples from that
receiver's neighbors.

The algorithm is very simple. In one trial, choose a uniformly random receiver $v$
and generate $K$ capped proposals from each of its
neighbors. Let $N_v$ be the number of these proposals
that land at $v$, and return $\min\{K,N_v\}/K$.
The expectation of this quantity, multiplied by $n/2$,
is the expected value of the random fractional matching
constructed above, averaging over the proposal samples.
Note that the estimator does not need proposal IDs or decisions about which incoming
copies are kept.

We return the average of $O(\eps^{-2}\log(2/\zeta))$ trials. Concentration of
independent bounded variables, together with the small
expected loss in the fractional construction, gives
additive error $\eps n$ with probability at least
$1-\zeta$. The total work remains
$(2\Delta)^{O(1/\eps)}\log(2/\zeta)$.
This estimates $\nu_f(G)$ on every graph and $\nu(G)$
on bipartite graphs. 

The estimator in \Cref{thm:deterministic} follows directly:
sample vertices uniformly, evaluate their total incident
fractional weight using the deterministic oracle, and
multiply the average by $n/2$. Only the choice of sampled
vertices is random.

\paragraph{AI assistance.}
AI tools contributed to the development of the
initial proof and subsequent refinements of the results,
as detailed in the AI methodology statement at the end
of the paper.

\subsection{Broader context}\label{sec:related}

\paragraph{Continuous optimization for matching and flows.}
Closest to our optimization approach, \citet{AJJST22}
use a box--simplex formulation and area-convex optimization
to obtain a deterministic semi-streaming
$(1-\eps)$-approximation for bipartite matching in
$O(\eps^{-1}\log(1/\eps)\log n)$ passes and $O(n)$
space. Their space bound improves on the earlier
linear-programming approach of \citet{AG18}.

Our optimizer builds on area convexity, introduced by
\citet{She17} for box--simplex games and approximate
undirected multicommodity flow. Subsequent applications
include optimal transport \citep{JST19}, mixed packing
and covering \citep{BSW19}, and dynamic bipartite matching
\citep{JJST22}. \citet{CST21} relate this approach to
relative Lipschitzness, and \citet{JT23} simplify the
optimization method by removing the high-accuracy
inner loop.
Our changes to the formulation and regularizer support
local updates and exact recursive sampling;
\Cref{rem:ajjst,rem:regularizer} discuss these choices.

\paragraph{Sampling and local simulation.}
Sampling can provide access to an implicitly represented
object without constructing it explicitly. Early examples
include reachability sketches \citep{Coh97} and sampling
methods for finding large dot products \citep{CL99}.
Related ideas accelerate approximate matrix multiplication
\citep{DKM06} and sublinear optimization \citep{CHW12}.

Sampling has also been used to simulate iterative local
algorithms without exploring their full dependency
neighborhoods. \citet{KMNT20} recursively sample a
fractional peeling process to estimate matching size from
uniform edge samples, and obtain a constant-factor matching
LCA using $O(\Delta\log n)$ queries.
For maximal independent set, \citet{Gha22} develops an
efficient local simulation of a randomized marking process,
obtaining an LCA with
$\operatorname{poly}(\Delta)\log n$ query complexity.

Our sampler also explores the graph adaptively, choosing
subsequent probes based on earlier outcomes.
Such adaptivity can be essential for efficiency:
\citet{ABGS25} show that non-adaptive LCAs require
$\Delta^{\Omega(\log\Delta/\log\log\Delta)}$ queries
even for constant-factor approximations of maximum matching.

\paragraph{Bernoulli factories.}
Bernoulli factories provide exact sampling from
distributions whose parameters are available only through
samples. They have been used to sample exponential-weight
distributions in mechanism design and entropy-regularized
matching \citep{DHKN21}, to round fractional points while
preserving their marginals \citep{NPS21}, and to resample
particles whose weights cannot be evaluated explicitly
\citep{SDD19}.

Here we use factories to implement the smooth update
formulas of a deterministic local optimization method.
The samplers and coins for each round use an expected
constant number of calls to those of the preceding round.
This gives exact access to the optimizer's distributions
with $2^{O(T)}$ expected work after $T$ rounds, without
numerically evaluating the iterates or accumulating
estimation error through the recursion.

\section{Preliminaries}

\subsection{Query and computation model}\label{sec:model}
The input is a finite simple graph $G=(V,E)$ with vertices
labeled by $[n]$, known $n=|V|$, and a known integer
$\Delta\ge2$ bounding the maximum degree.
A \emph{degree query} returns $\deg(w)$; a \emph{neighbor
query} returns the $i$-th neighbor of $w$ in a fixed
arbitrary order, for $1\le i\le\deg(w)$.
Each costs one graph query.
An algorithm may also sample a uniformly random vertex.
The query complexity counts only graph queries.
For $n\le1$, the answers are trivial.

\paragraph{Computational work.}
Numerical computations use finite binary representations,
with precision chosen as needed.
We charge for the bit operations used in arithmetic,
comparisons, and numerical evaluations of elementary
functions. Each graph query, random-bit draw, and uniform
vertex sample costs one unit of work.
Operations on vertex identifiers and the known value $n$
have unit machine-word cost.
In particular, elementary-function evaluations and coins
whose biases are specified by formulas are implemented
by finite procedures whose numerical work is charged.

The optimizer and sampler are specified by real-valued
formulas. \Cref{sec:finite-precision} gives finite
implementations that preserve the stated guarantees
and asymptotic work bounds.
All work bounds for the final LCAs and estimators refer
to these implementations.
The final algorithms terminate within their worst-case
query and work budgets on every execution.
The uncapped sampler of \Cref{sec:sampling} terminates
almost surely and has expected query and work bounds.

For the initial sampler analysis in \Cref{sec:sampling}
and the factory proofs in \Cref{sec:factory-proof},
we use an \emph{auxiliary operation count} that charges
one unit for each graph query, scalar arithmetic operation,
comparison, integer rounding, elementary-function evaluation,
and ideal random draw, including a known-bias coin.
This is an intermediate measure used in the analysis.
The final computational-work bounds use the
finite-precision convention above.

\paragraph{Randomness.}
Randomness is supplied by an oracle.
The estimator uses fresh independent randomness.
The LCA instead gives each proposal its own random seed,
which is reused whenever that proposal is needed
(\Cref{sec:proposal-access}).
A seed identifies a stream of independent unbiased bits.
A proposal is drawn by a recursive procedure that may
visit many vertices. All of its coins, at every vertex
it visits and in every recursive call, use fresh bits
from the proposal's own stream.
The outcome is therefore a deterministic function of
its seed, and every query that needs the proposal
recomputes the same outcome.
Different proposals have independent seeds, and their
computations need not agree on the random choices made
at vertices they both visit.

Rounding decisions are likewise reproducible:
each original proposal copy has an independent seed
for its marks across levels, reused by every query.
Seeds supply as many bits as needed; we neither bound
their length nor charge for their storage or addressing.

\subsection{Tools}
 
\begin{theorem}[\citet{Kon31}; see also \citet{Ege31}]\label{thm:konig}
In a bipartite graph the maximum size of a matching equals the minimum size of a vertex cover.
\end{theorem}
 
A \emph{fractional matching} of $G$ is a vector $f\ge0$ on $E$ with $\sum_{e\ni v}f_e\le1$ at every vertex $v$, and the \emph{fractional matching number} $\nu_f(G)$ is the largest value $\sum_ef_e$. A matching is a fractional matching, so $\nu_f(G)\ge\nu(G)$.
 
\begin{corollary}\label{cor:fractional}
Let $G$ be bipartite and let $f$ be a fractional matching of $G$. Then $\sum_e f_e\le\nu(G)$. Hence $\nu_f(G)=\nu(G)$.
\end{corollary}
\begin{proof}
Let $C$ be a minimum vertex cover, so $|C|=\nu(G)$ by \Cref{thm:konig}. Every edge has an endpoint in $C$, so $\sum_e f_e\le\sum_{v\in C}\sum_{e\ni v}f_e\le|C|$.
\end{proof}
 
The \emph{bipartite double cover} $B(G)$ of a graph $G$ has vertex set $V\times\{1,2\}$ and, for each edge $uv$ of $G$, the two edges $(u,1)(v,2)$ and $(v,1)(u,2)$. It reduces fractional matching in $G$ to matching in a bipartite graph.
 
\begin{lemma}[{Double cover; cf.\ \citep[Lemma~7]{AKO18} and \citep{DS24}}]\label{lem:double}
For every graph $G$:
\begin{enumerate}[label=(\alph*)]
\item $\nu(B(G))=2\nu_f(G)$;
\item every matching $M$ of $B(G)$ induces the half-integral fractional matching
\[
x_{uv}=\tfrac12\bigl(\mathbf 1[(u,1)(v,2)\in M]+\mathbf 1[(v,1)(u,2)\in M]\bigr)
\]
of $G$, of value $|M|/2$;
\item $B(G)$ has $2n$ vertices and maximum degree at most $\Delta$, and each degree or neighbor query to $B(G)$ is answered by one query to $G$.
\end{enumerate}
\end{lemma}
\begin{proof}
(b) At each vertex $u$ of $G$, the copy $(u,1)$ contributes at most $\frac12$ and the copy $(u,2)$ at most $\frac12$.
 
(a) A fractional matching $f$ of $G$ gives the fractional matching of $B(G)$ with weight $f_{uv}$ on both $(u,1)(v,2)$ and $(v,1)(u,2)$, of value $2\sum_ef_e$; since $B(G)$ is bipartite, \Cref{cor:fractional} gives $\nu(B(G))\ge2\nu_f(G)$. Part (b) gives the reverse inequality.
 
(c) The $i$-th neighbor of $(u,j)$ is $(w,3-j)$, where $w$ is the $i$-th neighbor of $u$.
\end{proof}
 
\begin{lemma}[Nonisolated vertices]\label{lem:active}
If $G$ has $n_0$ nonisolated vertices and maximum degree at
most $\Delta$, then
\[
n_0\le2\Delta\nu_f(G),\qquad |E|\le2\Delta\nu_f(G).
\]
\end{lemma}
\begin{proof}
The endpoints of a maximal matching form a vertex cover.
Every nonisolated vertex has a neighbor in this cover, and
every edge meets it. The sum of the degrees of the cover
vertices is at most $2\Delta$ times the matching size,
which is at most $2\Delta\nu_f(G)$.
\end{proof}

\begin{theorem}[Bounded differences; \citet{McD89}]\label{thm:mcd}
Let $Y_1,\dots,Y_m$ be independent and $W=f(Y_1,\dots,Y_m)$, where changing only $Y_i$ changes $f$ by at most $c_i$. If $V=\sum_i c_i^2>0$, then for every $t>0$,
\[
\Prob\{W-\E W\ge t\}\le e^{-2t^2/V},\qquad \Prob\{W-\E W\le -t\}\le e^{-2t^2/V}.
\]
In particular (Hoeffding's inequality \citep{Hoe63}), independent $Z_1,\dots,Z_m\in[0,1]$ satisfy $\Prob\{|\tfrac1m\sum_i(Z_i-\E Z_i)|>a\}\le 2e^{-2ma^2}$.
\end{theorem}

\section{Matching as a saddle-point problem}\label{sec:game}

We formulate fractional matching as a min--max problem with two
sets of variables: proposal distributions $q$, which maximize an
objective $F(x,q)$, and prices $x$, which minimize it. The objective
is affine in either set of variables when the other is fixed.
A solution is a \emph{saddle point} $(x^*,q^*)$, satisfying
\[
F(x^*,q)\leq F(x^*,q^*)\leq F(x,q^*)
\qquad\text{for every feasible }x,q.
\]
Thus, with prices fixed at $x^*$, no change to the proposal distributions
can increase the objective; with these distributions fixed at $q^*$,
no change to the prices can decrease it.

\paragraph{Variables.}
For each vertex $u$, let $I_u=N(u)\cup\{\dmy_u\}$, where $\dmy_u$
is a private \emph{dummy} target. Every vertex $u$ holds two variables:
\begin{itemize}
\item a \emph{proposal distribution}
$q_u=(q_{ua})_{a\in I_u}$, a probability vector on $I_u$:
vertex $u$ puts mass $q_{uv}$ on each neighbor $v$ and the remaining
mass $q_{u\dmy}$ on its dummy;
\item a \emph{price} $x_u\in[-1,1]$, associated with the proposal
mass that $u$ receives.
\end{itemize}
Terminology wise, we will refer to a draw from a proposal distribution $q_u$ (a neighbor of $u$ or the dummy) as a \emph{proposal}. Note that
the coordinates $q_{uv}$ and $q_{vu}$ are distinct, representing
proposal mass sent in opposite directions.

Let $I=\{(u,a):u\in V,\ a\in I_u\}$. Combining the proposal
distributions of vertices gives a vector $q\in\R^I$, and combining the prices
gives a vector $x\in\R^V$. Their feasible sets are
\[
Q=\left\{q\in\R^I_{\ge0}:
  \sum_{a\in I_u}q_{ua}=1\text{ for every }u\in V\right\},
\qquad
X=[-1,1]^V.
\]
Thus, $Q$ is a product of simplices, one for each vertex.
We write $Z=X\times Q$ for the joint feasible set.

\paragraph{Loads and accepted mass.} The \emph{load} of $v$ is the total proposal mass it receives,
\begin{equation} \label{eq:load}
    \ell_v(q)=\sum_{u\in N(v)}q_{uv}.
\end{equation}
Each vertex can accept at most one unit, so the total \emph{accepted mass}
is the sum of the loads capped at one,
\[
P(q)=\sum_{v\in V}\min\{1,\ell_v(q)\}.
\]
If $f$ is a fractional matching and every vertex $u$ assigns mass $f_{uv}$
to each neighbor $v$, with the rest on its dummy, then every load is at
most one and $P(q)=2\sum_e f_e$, because each edge's weight is counted
once from each endpoint. Hence the largest accepted mass is at least
$2\nu_f(G)$. \Cref{lem:value} shows that it is exactly $2\nu_f(G)$.

\paragraph{Prices.} Our goal is to maximize the accepted mass $P(q)$.
We introduce prices to express this optimization as a saddle-point problem
to which we can apply the area-convex method of
\Crefrange{sec:reg}{sec:opt} \citep{She17,JT23}.
\Cref{sec:sampling} then samples the resulting proposal distributions locally.
For a load $\ell$, the capped load $\min\{1,\ell\}$ is concave and nonsmooth.
The following identity expresses it, up to scaling and an additive
constant, as the minimum of affine functions:
\begin{equation}\label{eq:variational}
2\min\{1,\ell\}-1=\min_{x\in[-1,1]}\bigl[x+\ell(1-x)\bigr].
\end{equation}
The minimizing price is $+1$ when the vertex is overloaded ($\ell>1$)
and $-1$ when it is underloaded ($\ell<1$); when $\ell=1$, every price
in $[-1,1]$ is a minimizer.

\paragraph{The objective.} Applying~\eqref{eq:variational} at every vertex
$v$, with load $\ell_v(q)$ and price $x_v$, gives
\begin{equation}\label{eq:F}
F(x,q)=\sum_{v\in V}\bigl[x_v+\ell_v(q)(1-x_v)\bigr].
\end{equation}
Thus, $\min_{x\in X}F(x,q)=2P(q)-n$.
Maximizing this quantity over the proposal distributions is equivalent
to maximizing the accepted mass, so our target is
\[
\max_{q\in Q}\min_{x\in X}F(x,q)
=2\max_{q\in Q}P(q)-n=4\nu_f(G)-n.
\]
The following lemma verifies this value and shows that the minimum and
maximum can also be interchanged.

\begin{lemma}[Saddle-point value]\label{lem:value}
For every graph $G$:
\begin{enumerate}[label=(\alph*)]
\item $P(q)\le 2\nu_f(G)$ for every $q\in Q$.
\item $\min_{x\in X}F(x,q)=2P(q)-n$ for every $q\in Q$.
\item There is $q^*\in Q$ with $F(x,q^*)\ge 4\nu_f(G)-n$ for every $x\in X$, in which each vertex puts all of its mass on a single neighbor or on its dummy.
\item
$\displaystyle
\min_{x\in X}\max_{q\in Q}F(x,q)
=
\max_{q\in Q}\min_{x\in X}F(x,q)
=
4\nu_f(G)-n$.
\end{enumerate}
\end{lemma}
\begin{proof}
(a) For each ordered pair with $uv\in E$ put $f_{uv}=q_{uv}/\max\{1,\ell_v(q)\}$. Then $\sum_u f_{uv}=\min\{1,\ell_v(q)\}\le1$ and $\sum_v f_{uv}\le\sum_v q_{uv}\le1$. The symmetric weights $w_{\{u,v\}}=(f_{uv}+f_{vu})/2$ therefore form a fractional matching of $G$, of total weight $\frac12\sum f=P(q)/2$. Hence $P(q)/2\le\nu_f(G)$.
 
(b) Apply~\eqref{eq:variational} at each vertex; the coordinates of $x$ separate.
 
(c) Fix a maximum matching $M^*$ of $B(G)$, so $|M^*|=2\nu_f$ by \Cref{lem:double}(a). Let $u$ put unit mass on $v$ if $(u,1)(v,2)\in M^*$, and on its dummy if $(u,1)$ is unmatched. Then $\ell_v(q^*)=1$ if $(v,2)$ is matched and $0$ otherwise, so $\ell_v(q^*)=1$ at exactly $2\nu_f$ vertices and $F(x,q^*)=2\nu_f+\sum_{v:\ell_v(q^*)=0}x_v\ge 2\nu_f-(n-2\nu_f)$. For bipartite $G$ one can choose $M^*$ to consist of both copies of each edge of a maximum matching of $G$; then each matched vertex assigns unit mass to its partner.

(d) Parts (a)--(c) give
$\max_q\min_x F(x,q)=4\nu_f-n$.
Moreover, $F(x,q^*)\ge 4\nu_f-n$ for every $x$,
so $\min_x\max_q F(x,q)\ge 4\nu_f-n$.

For the reverse inequality, let $C$ be a minimum vertex cover
of $B(G)$, and write
$C_L=\{u:(u,1)\in C\}$ and $C_R=\{v:(v,2)\in C\}$.
Set $x_v=1$ for $v\in C_R$ and $x_v=-1$ otherwise.
If $u\notin C_L$, all its neighbors belong to $C_R$,
so its proposal mass contributes zero to
$\sum_{u,v}q_{uv}(1-x_v)$.
Each $u\in C_L$ contributes at most two.
Thus, for every $q\in Q$,
\[
F(x,q)\le 2|C_R|-n+2|C_L|
=2|C|-n=4\nu_f-n.
\]
\end{proof}

\begin{remark}[The double cover]\label{rem:double}
Each real coordinate $q_{uv}$ corresponds to the edge
$(u,1)(v,2)$ of the double cover $B(G)$. The proposal distributions
enforce the capacities on the proposer side $V\times\{1\}$, while
the prices represent the capacity constraints on the receiver side
$V\times\{2\}$.
\end{remark}

\begin{remark}[Optimization over prices]\label{rem:cover}
For fixed prices, the term $\sum_v x_v$ in $F(x,q)$ is constant.
Each unit of proposal mass sent to a neighbor $v$ contributes
$1-x_v$ to the objective, whereas mass sent to the dummy
contributes zero. Thus, $u$ can maximize its contribution by
assigning all its mass to the lowest-priced options, treating
the dummy as an option of price $1$.

Maximizing independently at each vertex gives a convex,
generally nonsmooth function of the prices alone,
\begin{equation}\label{eq:Fbar}
\overline F(x)=\max_{q\in Q}F(x,q)
=\sum_{v\in V}x_v
+\sum_{u\in V}\max\Bigl\{0,\ \max_{v\in N(u)}(1-x_v)\Bigr\}.
\end{equation}
To interpret this function, set $\lambda_v=(1+x_v)/2$.
These are the nonnegative Lagrange multipliers for the receiver-capacity
constraints $\ell_v(q)\le1$.
If the receiver copy $(v,2)$ has cover weight $\lambda_v$, the least
feasible weight on the proposer copy $(u,1)$ is
$\max\{0,\max_{v\in N(u)}(1-\lambda_v)\}$.
Hence $(\overline F(x)+n)/2$ is the cost of the cheapest fractional
vertex cover of $B(G)$ with these receiver weights.
Minimizing $\overline F$ over $X$ therefore solves the fractional
vertex cover problem on $B(G)$.
\end{remark}

\begin{remark}[{Comparison with \citet{AJJST22}}]\label{rem:ajjst}
\citet{AJJST22} also formulate bipartite matching as a bilinear
saddle-point problem over a box and a simplex, a formulation
known as a \emph{box--simplex game}. Their application was for the semi-streaming model. Their simplex is a single simplex over the edges of $G$ plus a dummy edge, scaled by an estimate $M$ of $\nu(G)$ that a greedy pass provides, and their coupling matrix is the unsigned incidence matrix, so each edge loads both of its endpoints. That casting does not suit our purpose of local computation. The scale $M$ is a global estimate of the very quantity we want and anyhow the normalization of a single simplex is global: in every round, the load at a vertex depends on a normalizing constant summed over all edges. Our formulation allows our steps to normalize separately at each vertex, so one round depends only on a few hops. 
Moreover, their casting does not directly support our local rounding
in \Crefrange{sec:prop}{sec:lca}, which requires sampling from a
proposal distribution at each vertex. Our casting provides these
per-vertex distributions, and \Cref{sec:sampling} constructs their
local samplers.
\end{remark}

\section{An area-convex regularizer}\label{sec:reg}

We now develop the ingredients of an optimization step that updates a state that comprises prices $x$ and proposal distributions $q$. 
The update directions aim to decrease $F$ in the prices and increase it in the proposal distributions. To control the movement from the current state, we use a convex scalar-valued function called a \emph{regularizer}. Its associated \emph{Bregman divergence}, defined below, provides the penalty for changing the state.
We then derive an explicit formula
for the regularized price update. In \Cref{sec:opt}, we combine
this price update with multiplicative updates of the proposal
distributions to obtain the full iterative method.

We choose the regularizer to satisfy two requirements. First, we
require \emph{area convexity}, which controls the interaction between
price and distribution updates and allows us to apply the
extragradient methods of \citet{She17,JT23}. In \Cref{sec:opt}, these
give an $O(\delta^{-1}\log(2\Delta))$ iteration bound for accepted-mass
error $\delta\nu_f(G)$. The initialization is chosen so that
the potential is proportional to the optimum, which gives
this multiplicative guarantee. 
Second, we require explicit price-update formulas with uniform bounds that support exact simulation from samples in \Cref{sec:sampling}.
We next fix the notation, and then
define our regularizer and establish these two properties.

\subsection{Optimization notation}\label{sec:opt-notation}

We use the standard Euclidean inner products on $\R^V$ and $\R^I$.
On pairs, the inner product is the sum of those on the two components:
\[
\inner{(x,p)}{(y,h)}=\inner{x}{y}+\inner{p}{h}.
\]
Gradients are taken with respect to these inner products. We write
$\mathbf 1\in\R^V$ for the vector with every coordinate equal to one.
Products and division of vectors are coordinatewise; for example,
$(x/y)_v=x_v/y_v$ and $(q\bar q)_{ua}=q_{ua}\bar q_{ua}$.
Powers, logarithms and exponentials are also applied coordinatewise.

\paragraph{The load map.}
The \emph{load map} $L:\R^I\to\R^V$ sums the proposal mass
arriving at each vertex:
\[
(Lq)_v=\ell_v(q)=\sum_{u\in N(v)}q_{uv}.
\]
As a matrix, $L$ has a row for each vertex $v$ and a column for
each pair $(u,a)\in I$, with entry one when $a=v$ and zero otherwise.
Its transpose, also called its \emph{adjoint},
$L^*:\R^V\to\R^I$, copies each vertex's price to the coordinates
that send mass to it:
$(L^*x)_{uv}=x_v$ for $v\in N(u)$, and
$(L^*x)_{u\dmy}=0$. Thus,
\[
\inner{Lq}{x}=\inner{q}{L^*x}
=\sum_{u\in V}\sum_{v\in N(u)}q_{uv}x_v.
\]
Let $b=L^*\mathbf 1$, the indicator of the non-dummy coordinates.
Then the objective can be written as
\begin{equation}\label{eq:F-op}
F(x,q)=\inner{\mathbf 1}{x}+\inner{b-L^*x}{q}.
\end{equation}
This is a box--simplex game in the sense of \citet{JT23}, with
a product of simplices in place of a single simplex;
\Cref{rem:ajjst} compares this casting with that of \citet{AJJST22}.

\paragraph{Gradients and the optimality gap.}
The \emph{descent--ascent operator} stacks the gradient of $F$
in the prices and minus its gradient in the proposal distributions:
\begin{equation}\label{eq:g}
g(x,q)=\bigl(\nabla_xF,\ -\nabla_qF\bigr)
=\bigl(\mathbf 1-Lq,\ L^*x-b\bigr).
\end{equation}
The signs encode descent in the prices and ascent in the proposal
distributions. Price descent lowers the prices of underloaded vertices
and raises those of overloaded vertices, while distribution ascent
favors cheaper neighbors.

For $z=(x,q)$ and a comparison point $w=(y,h)$ in $Z$,
expanding \eqref{eq:F-op} and using adjointness gives
\begin{equation}\label{eq:gap}
\inner{g(z)}{z-w}=F(x,h)-F(y,q).
\end{equation}
Maximizing over $w=(y,h)\in Z$ gives the \emph{saddle-point gap}
\[
\max_{h\in Q}F(x,h)-\min_{y\in X}F(y,q).
\]
This is the sum of the improvement available by changing the proposal
distributions alone and the improvement available by changing the
prices alone. It is nonnegative and vanishes exactly at a saddle point.
The analysis in \Cref{sec:opt} bounds the sum of the inner products
in \eqref{eq:gap} over iterations by changes in a potential,
which telescope.

\subsection{The regularizer}\label{sec:regularizer}
We choose a regularizer $R_\alpha$.
We use the \emph{negative entropy}
\[
H(q)=\sum_{(u,a)\in I}q_{ua}\log q_{ua},
\]
with natural logarithms and the convention $0\log 0=0$.
For the prices, use shifted negative binary entropy:
\begin{equation}\label{eq:psi}
\psi(x)=\frac{(1+x)\log(1+x)+(1-x)\log(1-x)}2,
\qquad -1\le x\le1.
\end{equation}
Thus $\psi(0)=0$, $\psi(\pm1)=\log2$, and, in the interior,
\[
\psi'(x)=\atanh x=\frac12\log\frac{1+x}{1-x},
\qquad \psi''(x)=\frac1{1-x^2}.
\]
For $\alpha>0$, define $R_\alpha:Z\to\R$ by
\[
R_\alpha(x,q)
=2\sum_v(1+\ell_v(q))\,\psi(x_v)+\alpha H(q).
\]
The entropy term leads to multiplicative updates of proposal
probabilities: probabilities are reweighted and then normalized
at each vertex. The first term couples each price to its incoming
load. Its curvature in $x_v$ is larger when the load is larger
and as $x_v$ approaches either endpoint of $[-1,1]$.
The choice of $\psi$ also gives the smooth price response derived
in \Cref{sec:box}.

\paragraph{The update penalty.}
Let $Q^\circ$ denote the relative interior of $Q$: all proposal
probabilities are positive, including those of the dummies.
For an interior base $z=(x,q)\in(-1,1)^V\times Q^\circ$ and
$w\in Z$, the \emph{Bregman divergence} generated by $R_\alpha$ is
\[
D^\alpha_z(w)
=R_\alpha(w)-R_\alpha(z)-\inner{\nabla R_\alpha(z)}{w-z}.
\]
It measures how far $R_\alpha(w)$ lies above the tangent plane
at $z$. For the choices $\alpha\ge10$ used below,
\Cref{lem:area} establishes convexity, so $D^\alpha_z(w)\ge0$,
with equality when $w=z$.
For comparison, the regularizer $R(z)=\frac12\|z\|_2^2$ gives
the squared-distance penalty
$D_z(w)=\frac12\|w-z\|_2^2$ used in ordinary gradient descent.

The Bregman divergence generated by $H$ is the \emph{relative entropy}
\[
K_p(h)=\sum_{(u,a)\in I}h_{ua}\log\frac{h_{ua}}{p_{ua}}
\qquad(p\in Q^\circ,\ h\in Q),
\]
where a term with $h_{ua}=0$ is zero.
This is the sum of the per-vertex Kullback--Leibler (KL) divergences;
it is nonnegative and vanishes exactly when $h=p$.
Thus, the entropy contribution to $D^\alpha_z(y,h)$ is
$\alpha K_q(h)$, which measures changes in the proposal distributions
relative to their current probabilities.
In particular, for $\beta\ge0$,
\[
D^{\alpha+\beta}_z(w)
=D^\alpha_z(w)+\beta K_q(w^q),
\]
where $w^q$ is the $Q$-component of $w$.

The update subproblems are built from expressions of the form
\begin{equation}\label{eq:regularized-update}
\underbrace{\inner{v}{w-z}}_{\text{descent--ascent term}}
+
\underbrace{D^\alpha_z(w)}_{\text{penalty for changing the state}},
\end{equation}
where $v$ is a scaled descent--ascent vector, such as $\eta g(z)$.
The method of \Cref{sec:opt} handles these subproblems with a fixed
number of price responses and multiplicative distribution updates. The price response for fixed distribution is derived in \Cref{sec:box}.

Differentiation gives
\begin{equation}\label{eq:grad}
\nabla_x R_\alpha=2(1+Lq)\,\atanh x,
\qquad
\nabla_q R_\alpha
=2L^*\psi(x)+\alpha(1+\log q).
\end{equation}

\subsection{Area convexity}\label{sec:areaconvexity}

Area convexity relates the curvature of the regularizer to the
bilinear coupling in $F$. The form we will use in the convergence proof
is \eqref{eq:area}: it bounds the interaction of two updates by
their divergences from a common base.

\begin{definition}[{Area convexity \citep{She17}}]
A continuous convex $R$ on $Z$ is \emph{$\eta$-area convex}
for $g$ if, for all $z,z',z^+\in Z$ with centroid
$c=(z+z'+z^+)/3$,
\[
\eta\inner{g(z')-g(z)}{z'-z^+}
\le R(z)+R(z')+R(z^+)-3R(c).
\]
\end{definition}

\begin{lemma}[{Scalar area inequality; cf.\
\citep[Lemma~1]{JT23} and \citep[Lemma~4]{JST19}}]
\label{lem:scalar}
Let $r_8(x,q)=qx^2+8q\log q$ for $|x|\le1$, $q\ge0$.
Then $r_8$ is convex, and for any three points $(x_i,q_i)$
with centroid $(\bar x,\bar q)$,
\[
\sum_{i=1}^3 r_8(x_i,q_i)-3r_8(\bar x,\bar q)
\ge
\tfrac13\bigl|
(x_2-x_1)(q_3-q_1)-(q_2-q_1)(x_3-x_1)
\bigr|.
\]
\end{lemma}

\begin{proof}
Assume first $q_i>0$. Let $r_4(x,q)=qx^2+4q\log q$.
In a direction $(a,b)$ its Hessian gives
\[
2qa^2+4xab+4b^2/q\ge qa^2,
\]
because the difference is
\[
(\sqrt q\,a+2xb/\sqrt q)^2
+4(1-x^2)b^2/q\ge0.
\]
So $r_4$ and $r_8=r_4+4q\log q$ are convex.

Write $a_i=x_i-\bar x$, $b_i=q_i-\bar q$.
Since the $a_i$ and $b_i$ sum to zero, the left side of the
claimed inequality equals the sum of the Bregman divergences
of $r_8$ from the centroid to the three points.
Along the segment from the centroid to point $i$,
Taylor's formula and the Hessian bound give a divergence
of $r_4$ of at least
\[
a_i^2\int_0^1(1-t)((1-t)\bar q+tq_i)\,dt
\ge \bar q a_i^2/3.
\]
For $q\log q$, the second derivative $1/q$ and
$q\le\max(q_i,\bar q)\le3\bar q$ give a divergence of at least
$b_i^2/(6\bar q)$. Hence the left side is at least
\[
\frac{\bar q}{3}\sum_i a_i^2
+\frac{2}{3\bar q}\sum_i b_i^2
\ge
\frac{2\sqrt2}{3}
\Bigl(\sum_i a_i^2\sum_i b_i^2\Bigr)^{1/2}.
\]
Up to sign, the oriented area equals
$\sum_i a_i(b_{i+1}-b_{i-1})$ (indices mod 3).
Since $\sum_i b_i=0$,
\[
\sum_i(b_{i+1}-b_{i-1})^2=3\sum_i b_i^2.
\]
Thus, by Cauchy--Schwarz, the area is at most
$\sqrt3(\sum_i a_i^2\sum_i b_i^2)^{1/2}$.
As $2\sqrt2/3\ge\sqrt3/3$, the claim follows.
Continuity extends convexity and the inequality to $q_i=0$.
\end{proof}

The standard regularizer, which combines a quadratic price term with negative entropy, is area convex
\citep[Lemma~1]{JT23}. We extend this property to $R_\alpha$
by proving that its difference from the standard regularizer,
with entropy coefficient $8$, is convex for $\alpha\ge10$.
Adding a convex function preserves area convexity.
\begin{lemma}[Area convexity of $R_\alpha$]\label{lem:area}
For $\alpha\ge10$, $R_\alpha$ is convex and
$\tfrac13$-area convex for $g$.
Moreover, for every interior $z$ and all $z',z^+\in Z$,
\begin{equation}\label{eq:area}
\tfrac13\inner{g(z')-g(z)}{z'-z^+}
\le D^\alpha_z(z')+D^\alpha_z(z^+).
\end{equation}
\end{lemma}

\begin{proof}
For every ordered pair $(u,v)$ with $uv\in E$, apply
\Cref{lem:scalar} to the points $(x_v,q_{uv})$, and add
the nonnegative Jensen gaps of the dummy terms
$8q_{u\dmy}\log q_{u\dmy}$.
A direct expansion shows that
$\inner{g(z')-g(z)}{z'-z^+}$ equals, up to sign, the sum
over pairs of the oriented areas; the constant parts of $g$ cancel.
Hence
\[
r(x,q)=\sum_v\ell_v(q)x_v^2+8H(q)
\]
is convex and $\tfrac13$-area convex.
This part is \citet[Lemma~1]{JT23} with $|A|=L$;
they state it for a single simplex, but the argument is entrywise,
and we include it for completeness.
The rest of the proof, which handles the $\psi$ term, is new.

Next, put $h(x)=2\psi(x)-x^2$. Then
\[
h'(x)=2(\atanh x-x),\qquad
h''(x)=\frac{2x^2}{1-x^2},\qquad h'(x)^2\le2h''(x).
\]
For the last inequality, use
$0\le\atanh x-x\le x/\sqrt{1-x^2}$ for $0\le x<1$
and symmetry. The upper bound follows by integrating
$(\atanh x)'\le(x/\sqrt{1-x^2})'$.
So the Hessian
$\begin{psmallmatrix}qh''&h'\\h'&2/q\end{psmallmatrix}$
of $qh(x)+2q\log q$ is positive semidefinite.
Continuity extends this convexity to the boundary.
Therefore
\begin{align*}
R_\alpha-r
={}&2\sum_v\psi(x_v)
+\sum_{(u,v):\,uv\in E}
  \bigl[q_{uv}h(x_v)+2q_{uv}\log q_{uv}\bigr]\\
&+2\sum_u q_{u\dmy}\log q_{u\dmy}
+(\alpha-10)H(q)
\end{align*}
is convex.
Adding a convex function preserves area convexity by Jensen's
inequality, which proves the first claim.

Finally, for an interior base $z$ and
$c=(z+z'+z^+)/3$,
\begin{align*}
R_\alpha(z)+R_\alpha(z')+R_\alpha(z^+)-3R_\alpha(c)
&=D^\alpha_z(z')+D^\alpha_z(z^+)-3D^\alpha_z(c)\\
&\le D^\alpha_z(z')+D^\alpha_z(z^+).
\end{align*}
Together with area convexity, this gives \eqref{eq:area}.
\end{proof}

\subsection{The price update: a box response}\label{sec:box}

We derive the price update by minimizing
\eqref{eq:regularized-update} over prices while holding the
candidate proposal distributions fixed. Fix an interior base
state $z=(x,q)$, a vector
$v=(v^x,v^q)\in\R^V\times\R^I$, and candidate distributions
$h\in Q$. Writing $w=(y,h)$ and dropping terms independent
of $y$ gives the subproblem
\begin{equation}\label{eq:price-subproblem}
\min_{y\in X}\bigl\{\inner{v^x}{y}+D^\alpha_z(y,h)\bigr\},
\qquad \alpha\ge10.
\end{equation}
We call its minimizing price vector $y$ the \emph{box response},
since $X=[-1,1]^V$. The only $\alpha$-dependent term is
$\alpha K_q(h)$, which is constant in $y$, so the response
does not depend on $\alpha$ for fixed $z,v,h$.

The subroutine $\Resp_{z,v}(h)$ in \Cref{alg:resp} returns
the updated prices. Each stage of \Cref{sec:opt} calls it
before and after updating the proposal distributions.
The change of variables $\tau=\psi'(x)=\atanh x$ makes
the price optimality condition linear.

\begin{algorithm}[H]
\small
\DontPrintSemicolon
\caption{Box response $\Resp_{z,v}(h)$}
\label{alg:resp}
\KwIn{Interior base state $z=(x,q)$; vector $v=(v^x,v^q)$;
candidate distributions $h\in Q$.}
\KwOut{Updated prices $y$.}
$A\gets\frac{\mathbf 1+\ell(q)}{\mathbf 1+\ell(h)}$,
\quad
$B\gets-\frac{v^x}{2(\mathbf 1+\ell(h))}$
\tcp*{loads from \eqref{eq:load}}
$y\gets\tanh(A\atanh x+B)$\;
\KwRet{$y$}\;
\end{algorithm}

\begin{lemma}[Box response]\label{lem:box}
\Cref{alg:resp} returns the unique minimizer $y$ of
\eqref{eq:price-subproblem}. The minimizer lies in $(-1,1)^V$.
Writing $\tau=\atanh x$ and $\tau'=\atanh y$, its
optimality condition is
\begin{equation}\label{eq:tau}
(\mathbf 1+\ell(h))\,\tau'
=(\mathbf 1+\ell(q))\,\tau-\tfrac12v^x.
\end{equation}
Equivalently, $\tau'=A\tau+B$ and $y=\tanh\tau'$.
\end{lemma}
\begin{proof}
Up to a term independent of $y$, the objective is
\[
2\sum_{j\in V}(1+\ell_j(h))\psi(y_j)
+\inner{v^x-\nabla_xR_\alpha(z)}{y}.
\]
The coordinates separate. Since $\psi'=\atanh$ is strictly
increasing from $-\infty$ to $+\infty$ on $(-1,1)$, each
coordinate has a unique interior minimizer. Its first-order
condition, using \eqref{eq:grad}, is \eqref{eq:tau}.
Solving for $\tau'$ and applying $\tanh$ gives the algorithm.
\end{proof}

\paragraph{Uniform bounds.}
For the sampling implementation in \Cref{sec:sampling}, we need
uniform bounds on the coefficients $A,B$ in \Cref{alg:resp}.
The next lemma obtains these bounds when the relevant loads
are within constant factors of the base load.
\Cref{lem:ratios} will verify this condition for every call
to $\Resp$.

All price-gradient vectors used below have the form
\[
v^x=\lambda(\mathbf 1-Lk),
\qquad k\in Q^\circ,\quad \lambda\in\{1/3,1/6\}.
\]
Here $k$ denotes the proposal distributions used to evaluate
the gradient; these may differ from both the base distributions
$q$ and the candidate distributions $h$.
At a nonisolated vertex $j$, define
\begin{equation} \label{eq:vertexloadparams}
r=\frac{\ell_j(q)}{1+\ell_j(q)},\qquad
z_h=\frac{\ell_j(h)}{\ell_j(q)},\qquad
z_k=\frac{\ell_j(k)}{\ell_j(q)}.
\end{equation}
Thus $r\in(0,1)$ represents the base load by a bounded quantity,
while $z_h,z_k$ express the other loads relative to it.
In the following lemma, we suppress the vertex subscripts
on $A,B$.

\begin{lemma}[Box coefficients]\label{lem:coef}
The coefficients in \Cref{alg:resp} can be written as
\[
w=1-r+rz_h,\qquad
A=\frac1w,\qquad
B=\frac{\lambda\bigl(r(z_k+1)-1\bigr)}{2w}.
\]
If $z_h,z_k\in[e^{-1},e]$, then
\[
w,A\in[e^{-1},e],\qquad |B|<2.
\]
At an isolated vertex, $A=1$ and $B=-\lambda/2$.
\end{lemma}

\begin{proof}
Multiplying the numerators and denominators in the definitions
of $A$ and $B$ by
$1-r=1/(1+\ell_j(q))$ gives the stated formulas.
Since $w$ is a convex combination of $1$ and $z_h$,
both $w$ and $A=1/w$ lie in $[e^{-1},e]$.
Also,
\[
|r(z_k+1)-1|\le e,
\qquad
|B|\le e^2/6<2.
\]
At an isolated vertex all loads vanish, giving
$A=1$ and $B=-\lambda/2$, which satisfy the same
bounds on $A$ and $B$.

\end{proof}

\begin{remark}[Why not the standard regularizer]
\label{rem:regularizer}
The standard area-convex regularizer for box--simplex games
\citep{She17,JST19,JT23} specializes in our setting to
$\sum_v\ell_v(q)x_v^2+\alpha H(q)$.
For positive candidate load $\ell_v(h)$, its box response
is a quotient with denominator $2\ell_v(h)$, clipped to
$[-1,1]$ \citep[Algorithm~4]{JT23}.
This response is nonsmooth at the clipping points, and its
load denominator can be arbitrarily small or zero. These
features prevent a direct application of the smooth exact
sampling primitives in \Cref{sec:sampling}.

Our regularizer replaces $\ell_v(q)x_v^2$ by
$2\ell_v(q)\psi(x_v)$ and adds $2\sum_v\psi(x_v)$.
The choice of $\psi$ gives a smooth price response without
clipping, while the added term changes the load denominator
to $1+\ell_v(h)$, keeping the formula defined at zero load.
Under constant bounds on relative loads, \Cref{lem:coef}
also bounds the response coefficients. The sampler uses a
bounded representation of the transformed price to evaluate
this response, as described in \Cref{sec:formula-coins}.

Binary entropy also supports the multiplicative convergence
guarantee: with prices initialized near $-1$,
\Cref{lem:phi0} bounds the initial potential by
$O(\nu_f(G)\log(2\Delta))$.
\Cref{lem:area} establishes area convexity for the modified
regularizer. The locality of normalization comes separately
from our product of vertex simplices, as explained in
\Cref{rem:ajjst}.
\end{remark}

\section{The optimization method}\label{sec:opt}

We present the optimization algorithm that constructs proposal
distributions with nearly maximum accepted mass.
\Cref{alg:optimization} gives the complete procedure, and
\Cref{alg:iteration} specifies a single round.
At round $t$, the state consists of main proposal distributions
$q_t$, auxiliary proposal distributions $\bar q_t$, and prices
$x_t$.
Starting from a fixed initialization, the algorithm updates this state over $T$ rounds.

We run the optimizer only on nonisolated vertices. Isolated
vertices always return their dummy and have no effect on
accepted mass or fractional matching. Thus the state coordinates
and potential sums in this section omit isolated vertices;
the graph still has its original $n$ vertex labels. This
restriction requires only a local degree query, not knowledge
of the number $n_0$ of nonisolated vertices.

Each round has two stages. A gradient stage computes tentative
prices and main proposal distributions. An \emph{extragradient}
stage then evaluates the gradient at that tentative state and
uses it to form the next state. Each stage performs a fixed
number of price and proposal-distribution updates.
Our final output is the average of the tentative proposal
distributions over all $T$ rounds.

The regularizer of \Cref{sec:reg} supplies the Bregman penalties
used in these updates. Within each stage, we first compute a
provisional price response and use it to update the proposal
distributions by multiplicative reweighting and normalization.
We then compute the price response for the updated distributions.
The extragradient stage also updates the auxiliary distributions.
Every price response uses the state at the start of the round
as its Bregman base. The area-convex analysis controls the
interaction between the two stages \citep{She17,JT23}.

The optimization is deterministic given $G$. After describing the
algorithm and stating its convergence guarantee, we establish
bounds for the two stages of a round and combine them to prove
convergence. 

Each round is local in the sense that the updated state at a vertex depends only
on the previous-round state within six hops. With our
initialization, the state and output proposal distribution at
a vertex after $T$ rounds are determined by a neighborhood
of radius $O(T)$. Therefore, \Cref{alg:optimization} already gives a direct local algorithm for its output proposal distributions. Each round consists of a constant
number of updates involving neighboring states, so computing the final
$\widehat q_u$ at a specified vertex requires exploring only a
neighborhood of radius $O(T)$. This gives a worst-case deterministic query bound
of $\Delta^{O(T)}=\Delta^{O(\delta^{-1}\log\Delta)}$.

In \Cref{sec:sampling}, we show that for the more limited task of drawing one exact sample from $\widehat q_u$, without constructing the full state, we can reduce the cost to $\Delta^{O(1/\delta)}$ expected
queries. \Cref{sec:opt-sampling} prepares for
\Cref{sec:sampling} by bounding changes of the state between consecutive rounds.
In \Cref{sec:sampling}, we use these bounds to sample the proposal
distributions recursively across rounds. Generating a sample for
one round requires only an expected constant number of samples
and coin flips from the preceding round, avoiding exploration
of all neighbors.

\begin{algorithm}[H]
\small
\DontPrintSemicolon
\SetKwFunction{OptRound}{Round}
\caption{Computing proposal distributions with nearly maximum accepted mass}
\label{alg:optimization}
\KwIn{Graph $G$, degree bound $\Delta\ge2$, and number of rounds $T\ge1$.}
\KwOut{Proposal distributions $\widehat q\in Q$.}

\tcp{Initialization for the potential bound in \Cref{lem:phi0}}
Set the output at each isolated vertex to its dummy\;
$x_0\gets(-1+1/\Delta)\mathbf 1$ on nonisolated vertices\;
\ForEach{nonisolated $u\in V$}{
  $q_{0,uv}\gets 1/(2\Delta^2)$ for every $v\in N(u)$\;
  $q_{0,u\dmy}\gets 1-\deg(u)/(2\Delta^2)$\;
}
$\bar q_0\gets q_0$\;

\BlankLine
\For{$t=0,\ldots,T-1$}{
$(x_{t+1},q_{t+1},\bar q_{t+1},q'_t)\gets$
\OptRound{$x_t,q_t,\bar q_t$}
\tcp*{\Cref{alg:iteration}}
}
$\widehat q\gets \frac1T\sum_{t=0}^{T-1}q'_t$
  \tcp*{average the tentative distributions}
\KwRet{$\widehat q$}\;
\end{algorithm}

The round routine uses the normalization map $\cN$. For a positive
weight vector $r\in\R^I_{>0}$, define
\begin{equation}\label{eq:normalizer}
\bigl(\cN(r)\bigr)_{ua}
=\frac{r_{ua}}{\sum_{b\in I_u}r_{ub}}.
\end{equation}
Thus, normalization is performed separately at each vertex.

\begin{algorithm}[H]
\small
\DontPrintSemicolon
\SetKwComment{Stage}{}{}
\caption{\texttt{Round}: one optimization round}
\label{alg:iteration}
\KwIn{Current state $(x,q,\bar q)$.}
\KwOut{Next state $(x^+,q^+,\bar q^+)$ and tentative
proposal distributions $q'$.}

$\eta\gets 1/3$\;
$z\gets(x,q)$ \tcp*{fixed base for every price response}

\BlankLine
\textbf{Gradient stage}
\tcp*[r]{\Cref{lem:gradoracle}, $\alpha=\beta=16$}
\Indp
$v\gets\eta g(x,q)$\tcp*{See \eqref{eq:g}}
$x^*\gets\Resp_{z,v}(q)$
  \tcp*{provisional price response; \Cref{alg:resp}}
$q'\gets\cN\!\left(
  q\exp\!\left(-\frac{v^q+2L^*(\psi(x^*)-\psi(x))}{16}\right)
  \right)$
  \tcp*{update distributions; \eqref{eq:normalizer}}
$x'\gets\Resp_{z,v}(q')$
  \tcp*{price response for $q'$; \Cref{alg:resp}}
\Indm
\BlankLine
\textbf{Extragradient stage}
\tcp*[r]{\Cref{lem:egoracle}, $\alpha=32$, $\beta=64$}
\Indp
$\tilde v\gets\frac{\eta}{2}g(x',q')$\tcp*{See \eqref{eq:g}}
$x_a\gets\Resp_{z,\tilde v}(\bar q)$
  \tcp*{provisional price response; \Cref{alg:resp}}
$a\gets\frac{1}{64}
  \bigl(\tilde v^q+2L^*(\psi(x_a)-\psi(x))\bigr)$\;
$q^+\gets\cN\bigl(\sqrt{q\bar q}\,e^{-a}\bigr)$
  \tcp*{update main distributions; \eqref{eq:normalizer}}
$x^+\gets\Resp_{z,\tilde v}(q^+)$
  \tcp*{price response for $q^+$; \Cref{alg:resp}}
$\hat s\gets\frac{1}{64}
  \bigl(\tilde v^q+2L^*(\psi(x^+)-\psi(x))\bigr)$\;
$\bar q^+\gets
  \cN\bigl(\bar q\sqrt{q/q^+}\,e^{-\hat s}\bigr)$
  \tcp*{update auxiliary distributions; \eqref{eq:normalizer}}
\Indm
\BlankLine
\KwRet{$(x^+,q^+,\bar q^+,q')$}\;
\end{algorithm}

The initialization gives interior prices and positive proposal
probabilities. By \Cref{lem:box}, every price response remains
in $(-1,1)^V$, and multiplicative reweighting and normalization
preserve positive probabilities. Thus all subsequent updates
are well defined, with the main and auxiliary distributions
remaining in $Q^\circ$. The target saddle point may lie on the
boundary of $Z$: the regularizer is finite there, while its
steep derivatives keep the finite iterates in the interior.

\subsection{Convergence guarantee}\label{sec:guarantee}

The output of \Cref{alg:optimization} satisfies the
following bound.

\begin{theorem}[Convergence rate]\label{thm:rate}
Let $G$ be a graph of maximum degree at most $\Delta$, and
let $\widehat q$ be the output of \Cref{alg:optimization}
after $T$ rounds.
With $B_\Delta:=200+400\log(2\Delta)$,
\[
0\le 2\nu_f(G)-P(\widehat q)\le 3B_\Delta\nu_f(G)/T.
\]
\end{theorem}

Consequently, setting
\[
T=\left\lceil\frac{3B_\Delta}{\delta}\right\rceil
=O\bigl(\delta^{-1}\log(2\Delta)\bigr)
\]
gives accepted mass $P(\widehat q)\ge(2-\delta)\nu_f(G)$.
The proof is in \Cref{sec:rate-proof}.

The deterministic iteration also gives algorithms for fractional
matching by evaluating its updates directly. For $q\in Q$, define
\begin{equation}\label{eq:det-fractional}
f_{\{u,v\}}(q)
=\frac12\left(
\frac{q_{uv}}{\max\{1,\ell_v(q)\}}
+
\frac{q_{vu}}{\max\{1,\ell_u(q)\}}
\right).
\end{equation}
As in the proof of \Cref{lem:value}(a), these weights form a
fractional matching, with
\[
\sum_{e\in E}f_e(q)=\frac12 P(q).
\]

\begin{corollary}[Deterministic LOCAL algorithm]
\label{cor:fractional-local}
Let $G$ be a graph on $n$ vertices with maximum degree at most
a known $\Delta\ge2$, and let $0<\eps<1/2$.
There is a deterministic LOCAL algorithm that computes a
fractional matching of value at least $(1-\eps)\nu_f(G)$ in
\[
O\!\left(\frac{\log(2\Delta)}{\eps}\right)
\]
rounds. Each endpoint of an edge outputs the same edge weight.
\end{corollary}

\begin{proof}
Run \Cref{alg:optimization} for
$T=\lceil3B_\Delta/\eps\rceil$ rounds and let
$f^*=f(\widehat q)$.
By \Cref{thm:rate},
\[
\sum_e f^*_e
=\frac12P(\widehat q)
\ge (1-\eps/2)\nu_f(G).
\]
Each optimization round consists of a constant number of
local updates and exchanges between neighbors: loads are
incoming sums, proposal distributions are normalized at their
source, and price updates are computed at their vertex.
The final averaging and~\eqref{eq:det-fractional} add only
constant locality. Thus each $f^*_e$ depends on a neighborhood
of radius $O(T)$ around $e$.

Each vertex gathers the required neighborhood in $O(T)$ LOCAL
rounds. To obtain finite outputs, the endpoints use the same
deterministic procedure to compute a certified rational
approximation $\widetilde f_e$ satisfying
\[
0\le \widetilde f_e\le f^*_e,
\qquad
f^*_e-\widetilde f_e\le \eps/(4\Delta).
\]
Such approximations can be computed from the explicit update
formulas. Decreasing edge weights preserves feasibility, and
the total decrease is at most
$|E|\eps/(4\Delta)\le\eps\nu_f(G)/2$ by \Cref{lem:active}.
\end{proof}

\begin{corollary}[Deterministic fractional-matching LCA]
\label{cor:fractional-lca}
Under the same assumptions, there is a deterministic LCA
answering edge queries consistently with one fractional
matching of value at least $(1-\eps)\nu_f(G)$, using
\[
(2\Delta)^{O(\eps^{-1}\log(2\Delta))}
\]
graph queries and computational work per edge query,
in the model of \Cref{sec:model}.
Both bounds are worst case.
No preprocessing or persistent memory is required.
\end{corollary}
\begin{proof}
Use the finite implementation in
\Cref{prop:finite-optimizer}, with
$T=O(\eps^{-1}\log(2\Delta))$ optimization rounds.
Its output is a feasible fractional matching of value
at least $(1-\eps)\nu_f(G)$.
An edge query evaluates this construction within the
radius-$O(T)$ neighborhood on which its answer depends.
The precision and computational-work bounds in that
proposition give
$(2\Delta)^{O(T)}$ graph queries and computational work.
All answers use the same initialization, update rules,
and numerical rounding rules, so they describe one
fractional matching independently of the query order.
\end{proof}

For bipartite graphs, the algorithms of
\Cref{cor:fractional-local,cor:fractional-lca} can be combined
with deterministic rounding in the corresponding model to
obtain integral matchings. For example, combining
\Cref{cor:fractional-local} with
\citet[Theorem~3]{AKO18} gives a deterministic LOCAL algorithm
that computes a matching of size at least $(1-\eps)\nu(G)$,
splitting the error allowance between optimization and rounding, in
$O((\log^2(\Delta/\eps)+\log^* n)/\eps)$ rounds.

\subsection{Mirror descent and update inequalities}\label{sec:oracles}

For current proposal distributions $p\in Q^\circ$, a score vector $a$
and $\beta>0$, an \emph{entropic mirror-descent step} is
\[
p^+=\arg\min_{h\in Q}\bigl\{\inner{a}{h}+\beta K_p(h)\bigr\}
=\cN(p\,e^{-a/\beta})=:\MD^\beta_p(a).
\]
Thus the update favors smaller scores while penalizing deviation from
$p$. It multiplies probabilities by score-dependent factors and
normalizes separately at each vertex: this is the multiplicative-weights
form of mirror descent. The identity used in the analysis is, for all $h\in Q$,
\begin{equation}\label{eq:md}
\inner{a}{p^+-h}=\beta\bigl[K_p(h)-K_{p^+}(h)-K_p(p^+)\bigr],
\end{equation}
because $\log(p^+/p)=-a/\beta$ plus block constants, which vanish against $p^+-h$.

To combine this distribution update with a price update, fix an interior
base $z=(x,q)$ and a vector $v=(v^x,v^q)$
and $\alpha\ge10$. For each candidate $h$, minimize over the prices using
$y(h)=\Resp_{z,v}(h)$. The remaining regularized objective
and its score vector are
\[
f(h)=\inner{v}{(y(h),h)}+D^\alpha_z(y(h),h),\qquad a(h)=v^q+2L^*(\psi(y(h))-\psi(x))+\alpha\log(h/q).
\]

\begin{lemma}[{Partial minimization; cf.\ \citep[Lemmas~4--5]{JT23} and \citep{CST21}}]\label{lem:f}
Both $f$ and $\alpha H-f$ are convex, $\nabla f=a$ on $Q^\circ$ up to block constants, and for $p,p'\in Q^\circ$, $h\in Q$,
\begin{equation}\label{eq:smooth}
\inner{a(p')-a(p)}{p'-h}\le\alpha\bigl[K_p(p')+K_{p'}(h)\bigr].
\end{equation}
\end{lemma}
\begin{proof}
$f$ is a partial minimum of a jointly convex function (\Cref{lem:area}), hence convex. For fixed $y$ the minimized expression is $\alpha H(h)$ plus an affine function of $h$, so $f-\alpha H$ is an infimum of affine functions and is concave. By the envelope theorem and~\eqref{eq:grad}, $\nabla f(h)=a(h)$ up to block constants. Convexity of $f$ and of $\alpha H-f$ gives $0\le D^f_p(h)\le\alpha K_p(h)$. The three-point identity $\inner{\nabla f(p')-\nabla f(p)}{p'-h}=D^f_p(p')+D^f_{p'}(h)-D^f_p(h)$ then gives~\eqref{eq:smooth}; continuity covers boundary $h$.
\end{proof}

\begin{lemma}[{Gradient update; cf.\ \citep[Definition~4]{JT23}}]\label{lem:gradoracle}
Let $\beta\ge\alpha\ge10$, $q'=\MD^\beta_q(a(q))$, $x'=y(q')$ and $z'=(x',q')$. For every $w\in Z$,
\[
\inner{v}{z'-w}\le D^{\alpha+\beta}_z(w)-D^\alpha_{z'}(w)-D^\alpha_z(z').
\]
\end{lemma}
\begin{proof}
Add~\eqref{eq:md} for $a(q)$ to~\eqref{eq:smooth} with $(p,p',h)=(q,q',w^q)$; since $\beta\ge\alpha$, this gives $\inner{a(q')}{q'-w^q}\le\beta K_q(w^q)$. Stationarity of the box response gives $v^x+\nabla_xR_\alpha(z')-\nabla_xR_\alpha(z)=0$, and the $q$-component of $v+\nabla R_\alpha(z')-\nabla R_\alpha(z)$ is $a(q')$ up to block constants. So $\inner{v+\nabla R_\alpha(z')-\nabla R_\alpha(z)}{z'-w}\le\beta K_q(w^q)$. The three-point identity for $R_\alpha$ and $D^{\alpha+\beta}_z=D^\alpha_z+\beta K_q$ finish the proof.
\end{proof}

\begin{lemma}[{Extragradient update; cf.\ \citep[Definition~5]{JT23}}]\label{lem:egoracle}
Let $\beta\ge\alpha\ge10$ and $\bar q\in Q^\circ$. Set $q^+=\MD^\beta_{\bar q}(a(\bar q))$, $x^+=y(q^+)$, $z^+=(x^+,q^+)$ and $\bar q^+=\MD^\beta_{\bar q}(a(q^+))$. For every $w\in Z$,
\[
\inner{v}{z^+-w}\le D^\alpha_z(w)-D^\alpha_{z^+}(w)-D^\alpha_z(z^+)+\beta K_{\bar q}(w^q)-\beta K_{\bar q^+}(w^q).
\]
\end{lemma}
\begin{proof}
Apply~\eqref{eq:md} to the first update with comparator $\bar q^+$ and to the second with comparator $w^q$, and add:
\[
\inner{a(q^+)}{q^+-w^q}=\beta\bigl[K_{\bar q}(w^q)-K_{\bar q^+}(w^q)-K_{q^+}(\bar q^+)-K_{\bar q}(q^+)\bigr]+\inner{a(q^+)-a(\bar q)}{q^+-\bar q^+}.
\]
By~\eqref{eq:smooth} the last term is at most $\alpha[K_{\bar q}(q^+)+K_{q^+}(\bar q^+)]$, which is absorbed since $\beta\ge\alpha$. Conclude as in \Cref{lem:gradoracle}.
\end{proof}

In the extragradient stage of \Cref{alg:iteration}, $\alpha/\beta=1/2$.
The term $\alpha\log(h/q)$ in $a(h)$ therefore gives the square-root
factors in the two distribution updates.

\subsection{Convergence analysis}\label{sec:rate-proof}

Our proof adapts the potential argument of
\citet[Corollary~1]{JT23} to the regularizer $R_\alpha$ and
the product of vertex simplices $Q$. We use the update
inequalities established in \Cref{sec:oracles}, bound the
initial potential for our initialization, and translate the
resulting bound into an accepted-mass guarantee.

For each fixed comparison point $w$, we bound the accumulated quantity
$\sum_{t<T}\inner{g(z'_t)}{z'_t-w}$ by changes in a potential.
The identity \eqref{eq:gap} then translates this bound into an
accepted-mass guarantee. The potential combines divergences for the
main and auxiliary states.

For $z_t=(x_t,q_t)$ and $z'_t=(x'_t,q'_t)$ let $\Phi_t(w)=D^{32}_{z_t}(w)+64K_{\bar q_t}(w^q)$.

\begin{lemma}[{Per-round regret; cf.\ \citep[Lemma~2]{JT23}}]\label{lem:regret}
For all $w\in Z$ and all $t$, $\ \eta\inner{g(z'_t)}{z'_t-w}\le 2\Phi_t(w)-2\Phi_{t+1}(w)$.
\end{lemma}
\begin{proof}
Write $z=z_t$, $z'=z'_t$, $z^+=z_{t+1}$. \Cref{lem:gradoracle} with comparator $z^+$, and twice \Cref{lem:egoracle} with comparator $w$, give
\begin{align*}
\inner{\eta g(z)}{z'-z^+}&\le D^{32}_z(z^+)-D^{16}_{z'}(z^+)-D^{16}_z(z'),\\
\inner{\eta g(z')}{z^+-w}&\le 2\Phi_t(w)-2\Phi_{t+1}(w)-2D^{32}_z(z^+).
\end{align*}
Adding these and $\eta\inner{g(z')-g(z)}{z'-z^+}$ gives $\eta\inner{g(z')}{z'-w}$ on the left. The excess on the right is at most $\eta\inner{g(z')-g(z)}{z'-z^+}-D^{16}_z(z')-D^{32}_z(z^+)\le0$ by~\eqref{eq:area} and $D^{32}\ge D^{16}$.
\end{proof}

\begin{lemma}[Initial potential]\label{lem:phi0}
Let $q^*$ be the comparison distributions constructed in
\Cref{lem:value}(c). If $y\in\{-1,1\}^{V}$ has at most
$2\nu_f(G)$ positive coordinates among the nonisolated vertices,
then $\Phi_0(y,q^*)\le B_\Delta\nu_f(G)$.
\end{lemma}
\begin{proof}
Write $b=-1+1/\Delta$ and $\nu=\nu_f(G)$ within this proof.
The comparison loads lie in $\{0,1\}$ and sum to $2\nu$.
The initial total load satisfies
\[
\mu_0:=\sum_v\ell_v(q_0)=\frac{|E|}{\Delta^2}\le\nu,
\]
using \Cref{lem:active} and $\Delta\ge2$.
For the scalar Bregman divergence
$d_\psi(y,b)=\psi(y)-\psi(b)-\psi'(b)(y-b)$, direct
substitution gives
\[
d_\psi(-1,b)=-\log\!\left(1-\frac1{2\Delta}\right)
\le\frac1\Delta,\qquad
d_\psi(1,b)=\log(2\Delta).
\]
Thus most coordinates can be compared with $-1$ cheaply;
only the at most $2\nu$ positive coordinates pay a logarithmic
cost.

Expanding the potential, with sums over nonisolated vertices,
gives
\begin{align*}
\Phi_0(y,q^*)={}&
2\sum_v(1+\ell_v(q^*))d_\psi(y_v,b)\\
&+2\psi'(b)\sum_v(\ell_v(q^*)-\ell_v(q_0))(y_v-b)
+96K_{q_0}(q^*).
\end{align*}
The first term is at most $8\nu+8\nu\log(2\Delta)$:
use $n_0\le2\Delta\nu$ for negative coordinates and
$1+\ell_v(q^*)\le2$ for positive ones.
The absolute value of the second term is at most
$6\nu\log(2\Delta)$, since $|y_v-b|\le2$,
$2|\psi'(b)|=\log(2\Delta-1)$, and the two total loads
are at most $2\nu$ and $\nu$.

Finally, $q^*$ assigns mass one to $2\nu$ real coordinates;
the remaining mass is on dummies. Therefore
\[
K_{q_0}(q^*)
\le2\nu\log(2\Delta^2)
  +\sum_u-\log\!\left(1-\frac{\deg(u)}{2\Delta^2}\right)
\le2\nu\log(2\Delta^2)+2\nu.
\]
For the last step, use $-\log(1-a)\le2a$ for
$0\le a\le1/2$ and $\sum_u\deg(u)/\Delta^2=2\mu_0$.
Combining the bounds gives
$\Phi_0(y,q^*)\le[200+398\log(2\Delta)]\nu
\le B_\Delta\nu$.
\end{proof}

\begin{proof}[Proof of \Cref{thm:rate}]
Let $q^*$ be as in \Cref{lem:value}(c). Choose the minimizing
price $w^x_v=1$ when $\ell_v(\widehat q)>1$ and $w^x_v=-1$
otherwise, and put $w=(w^x,q^*)$. The number of positive
coordinates is at most $P(\widehat q)\le2\nu_f(G)$, so
\Cref{lem:phi0} applies.
Summing \Cref{lem:regret} over $t<T$, using $\Phi_T(w)\ge0$,
\eqref{eq:gap}, and the fact that $F$ is affine in $q$, gives
\[
\eta T\Bigl[
\frac1T\sum_{t<T}F(x'_t,q^*)-F(w^x,\widehat q)
\Bigr]\le 2\Phi_0(w).
\]
By \Cref{lem:value}(b,c), the bracket is at least
\[
(4\nu_f(G)-n_0)-(2P(\widehat q)-n_0)
=2\bigl(2\nu_f(G)-P(\widehat q)\bigr).
\]
Therefore, using $\eta=1/3$ and \Cref{lem:phi0},
\[
2\nu_f(G)-P(\widehat q)
\le \frac{\Phi_0(w)}{\eta T}
\le \frac{3B_\Delta\nu_f(G)}{T}.
\]
The lower bound follows from \Cref{lem:value}(a).
\end{proof}

\subsection{Properties used by the sampler}\label{sec:opt-sampling}

We next show that each round changes proposal probabilities and loads
by only constant factors. We also express the auxiliary
proposal distributions through ratios to the main distributions.
Together with \Cref{lem:coef}, these bounds allow the
sampler in \Cref{sec:sampling} to generate samples for
each new round using an expected constant number of
samples and coin flips from the preceding round.

We call the quantities negated in the exponential factors of the
proposal updates the \emph{scores}: a score $c$ gives a multiplier
$e^{-c}$. For the three proposal updates in \Cref{alg:iteration},
the scores on a real coordinate $(u,v)$ are
\begin{align*}
&\tfrac{1}{16}\bigl[\tfrac13(x_v-1)+2(\psi(x^*_v)-\psi(x_v))\bigr]
&&\text{(gradient stage)},\\
a_{uv}
&=\tfrac1{64}\bigl[\tfrac16(x'_v-1)+2(\psi(x_{a,v})-\psi(x_v))\bigr]
&&\text{(extragradient stage, main distributions)},\\
\hat s_{uv}
&=\tfrac1{64}\bigl[\tfrac16(x'_v-1)+2(\psi(x^+_v)-\psi(x_v))\bigr]
&&\text{(extragradient stage, auxiliary distributions)}.
\end{align*}
All three scores vanish on dummy coordinates.

For a vector $c\in\R^I$, define its \emph{oscillation} by
\[
\osc(c)=\max_{u\in V}
\left(\max_{a\in I_u}c_{ua}-\min_{a\in I_u}c_{ua}\right).
\]
This is the largest range of coordinate values within a single
vertex's block. Adding the same constant to every coordinate
in a block leaves the oscillation unchanged. This lets us ignore
the additive constants introduced by per-vertex normalization
when analyzing updates in logarithmic coordinates.

Define the coordinatewise ratio of the auxiliary to the main
proposal distributions by
\[
\rho_t=\bar q_t/q_t.
\]

\begin{lemma}[Bounded tilts]\label{lem:ratios}
For all $t$:
\begin{enumerate}[label=(\alph*)]
\item every coordinate of the scores satisfies
$|a|,|\hat s|<1/24$, hence
$|\tfrac32a-\hat s|<5/48$, and the gradient-stage score
lies in $(-1/6,1/8)$;
\item $\osc(\log\rho_t)\le 5/18$;
\item every coordinate of $\bar q_t/q_t$, $q'_t/q_t$,
$q_{t+1}/q_t$ and $\bar q_{t+1}/\bar q_t$ lies in
$[e^{-1},e]$, and so does the corresponding load ratio
at every nonisolated vertex.
\end{enumerate}
\end{lemma}

\begin{proof}
(a) All prices appearing in the updates lie in $(-1,1)$,
and their regularizer values lie in $[0,\log2]\subset[0,1]$.
The bounds follow from the score formulas above.

(b) In logarithmic coordinates, normalization adds a constant
within each vertex's block. Thus the main update gives
\[
\log(q_{t+1}/q_t)
=\tfrac12\log\rho_t-a+\text{block constants}.
\]
Substituting this into the auxiliary update gives
\[
\log\rho_{t+1}
=\tfrac14\log\rho_t+\tfrac32a-\hat s
+\text{block constants}.
\]
Oscillation ignores these constants. By (a),
$\osc(a),\osc(\hat s)<1/12$, so
\[
\osc(\log\rho_{t+1})
\le \tfrac14\osc(\log\rho_t)+\tfrac5{24}.
\]
Since $\rho_0=\mathbf 1$, induction gives
$\osc(\log\rho_t)\le5/18$.

(c) For $p,h\in Q^\circ$, each block of $\log(h/p)$
contains a nonnegative and a nonpositive coordinate,
because both distributions sum to one on that block.
Thus every coordinate has absolute value at most
$\osc(\log(h/p))$.

By (a), (b), and the update formulas, the relevant
oscillations satisfy
\begin{align*}
\osc(\log(\bar q_t/q_t))
&\le \tfrac5{18},\\
\osc(\log(q'_t/q_t))
&\le \tfrac7{24},\\
\osc(\log(q_{t+1}/q_t))
&\le \tfrac12\cdot\tfrac5{18}+\tfrac1{12}
 =\tfrac29,\\
\osc(\log(\bar q_{t+1}/\bar q_t))
&\le \tfrac12\cdot\tfrac29+\tfrac1{12}
 =\tfrac7{36}.
\end{align*}
All are below one, so exponentiating gives the coordinate
ratio bounds. Finally, at a nonisolated vertex $v$,
\[
\frac{\ell_v(h)}{\ell_v(p)}
=\sum_{u\in N(v)}
  \frac{p_{uv}}{\ell_v(p)}\,\frac{h_{uv}}{p_{uv}}
\]
is a weighted average of coordinate ratios, giving the
same bounds for the load ratios.
\end{proof}

In every box response of a round, $h\in\{q,q',\bar q,q^+\}$ and $k\in\{q,q'\}$, all relative to the same base $q$. Thus \Cref{lem:ratios}(c) verifies the hypotheses of \Cref{lem:coef} for every response.

\begin{proposition}[Reduced auxiliary update]\label{prop:reduced}
Let $\rho=\bar q/q$. Then $q^+=\cN(q\sqrt\rho\,e^{-a})$ and $\bar q^+=q^+\rho^+$, where
\[
W=\rho^{1/4}e^{\frac32a-\hat s},\qquad \rho^+_{ua}=\frac{W_{ua}}{\sum_{b\in I_u}q^+_{ub}W_{ub}}.
\]
\end{proposition}
\begin{proof}
On one block let $Z=\sum_b q_b\sqrt{\rho_b}e^{-a_b}$, so $U_a:=q^+_a/q_a=\sqrt{\rho_a}e^{-a_a}/Z$. The auxiliary update divided by $q^+_a$ is proportional to $\rho_aU_a^{-3/2}e^{-\hat s_a}=Z^{3/2}W_a$. A ratio of two distributions has mean one under the denominator, which fixes the normalization.
\end{proof}

Thus the sampler can represent the auxiliary proposal distributions by the ratio $\rho_t$ to the main proposal distributions $q_t$, without a separate sampler for $\bar q_t$.

\section{Exact local sampling}\label{sec:sampling}

We now implement a recursive sampler from the proposal distributions of \Cref{sec:opt}.
Given a vertex $u$, the sampler returns
one exact draw from the averaged distribution $\widehat q_u$,
without computing that distribution or the optimization state.

The sampling algorithm follows a simple rule: draw from the
previous proposal distribution and correct the draw by rejection
sampling. The bounds of \Cref{sec:opt-sampling} ensure that each
trial succeeds with probability at least an absolute constant.
The additional ingredient that we need is a way to make the acceptance decision
without computing its probability.

To see the idea, consider a bounded reweighting of $q$ to obtain $h$ at vertex $u$:
\begin{equation}\label{eq:sampling-tilt}
h_{ua}=\frac{q_{ua}A_{ua}}{Z_u},\qquad
Z_u=\sum_{b\in I_u}q_{ub}A_{ub},\qquad
\kappa\le A_{ua}\le1,
\end{equation}
where $\kappa>0$ is an absolute constant. If the weights were
available numerically, we could sample $h_u$ as follows:
\begin{enumerate}[label=(\arabic*)]
\item Draw $a\sim q_u$ and an independent $U\sim\mathrm{Unif}[0,1]$.
\item Return $a$ if $U\le A_{ua}$; otherwise repeat.
\end{enumerate}

A trial accepts and returns $a$ with probability
$q_{ua}A_{ua}$, so the accepted draw has law $h_u$.
The acceptance probability is $Z_u\ge\kappa$, giving at
most $1/\kappa$ trials in expectation. There is no need
to compute the normalizer $Z_u$.

Computing the weight $A_{ua}$ exactly is the obstacle:
Our update formulas depend on prices and loads, and
evaluating even one load
$\ell_v(q)=\sum_{u\in N(v)}q_{uv}$ can require a recursive
evaluation for every neighbor. This can introduce
$\Delta$-fold branching, whereas we seek constant expected
branching in our recursive sampler.

We replace the acceptance test in step (2) by a procedure
that returns one with probability exactly $A_{ua}$, using
fresh flips of previously available coins. Bernoulli
factories (\Cref{sec:acceptance-primitive}) let us construct
this acceptance coin with an expected constant number of
input flips, without computing its bias.

Our sampler follows the optimization round by round,
representing distributions by samplers and numerical
quantities by coins. Assume recursively that the preceding
round's samplers and coins are available. We implement the
scalar update formulas by Bernoulli factories, which construct
new coin procedures from the available ones. For the
reweighting in \eqref{eq:sampling-tilt}, we combine a draw
$a\sim q_u$ with the newly constructed acceptance coin,
repeating with fresh randomness until acceptance. The
resulting sampler for $h_u$ is then available for subsequent
updates.

Each input-coin flip may itself invoke earlier coins or
samplers. We show below that producing one sample or coin
flip for a new round uses an expected constant number of
calls to the preceding round's procedures. This yields the constant expected branching that we seek. The total cost
is obtained by expanding these calls recursively.
\Cref{sec:sampler-round} specifies the sampler;
\Cref{sec:sampler-exact} proves its exactness, and
\Cref{sec:sampler-cost} bounds its query cost and
arithmetic work.

\subsection{Bernoulli factories}\label{sec:acceptance-primitive}

A \emph{$p$-coin} is a procedure returning an independent
$\mathrm{Bern}(p)$ bit on each call. A \emph{Bernoulli factory}
for a function $f$ constructs an $f(p_1,\ldots,p_k)$-coin
using independent flips of input $p_1,\ldots,p_k$-coins and
auxiliary randomness, without knowing the input biases.

The next two lemmas give factory constructions for two
classes of functions: fixed elementary formulas and smooth
functions of one variable. Under their respective assumptions,
each construction uses an expected constant number of input
flips and expected constant additional work.\footnote{
Here and in the initial sampler analysis, work refers to
the auxiliary operation count defined in \Cref{sec:model}.
The finite-precision implementation is analyzed in
\Cref{sec:finite-precision}.}
The constructions simulate the acceptance decision directly,
without estimating the unknown biases of the input coins.

The elementary-formula construction combines the linear
Bernoulli factory of \citet{Huber16}, the fractional-power
factory of \citet{Mendo19}, and operations on bounded
unbiased random variables.
Both proofs are in \Cref{sec:factory-proof}.

\begin{lemma}[Elementary-formula factories]\label{lem:factory}
Let $\Phi(p_1,\ldots,p_k)$ be a fixed finite formula built from
input coordinates and real constants using addition,
subtraction, multiplication, division, exponentials, square
roots, and fourth roots. Let $S\subseteq[0,1]^k$ be a set of
valid inputs. Suppose that, for known constants $\beta>0$ and
$0<\gamma<1/2$, every $p\in S$ satisfies:
\begin{enumerate}[label=(\alph*)]
\item every denominator and every root argument in the formula
is at least $\beta$;
\item $\Phi(p)\in[\gamma,1-\gamma]$.
\end{enumerate}
Given independent coins of biases $p_1,\ldots,p_k$, with
$p\in S$, there is an exact $\Phi(p)$-coin that terminates
almost surely. Its expected number of input flips and
expected additional work are bounded by constants depending
only on $\Phi,\beta,\gamma$, uniformly over $p\in S$.
\end{lemma}

\begin{lemma}[{Smooth one-variable factory; cf.\ \citep{NP05,HNP11}}]
\label{lem:smooth-factory}
Fix $0<\gamma<1/2$ and a function
$f:[0,1]\to[\gamma,1-\gamma]$ with four continuous derivatives
on $[0,1]$ (a $C^4$ function). Suppose bounds on these
derivatives are known, and $f$ and $f''$ have fixed formulas
that can be evaluated in constant work at rational points.
Given independent flips of a $p$-coin, there is an exact
$f(p)$-coin that terminates almost surely and uses an expected
constant number of input flips and expected constant additional
work. These constants depend only on $f$ and $\gamma$, not on
$p$. 
\end{lemma}

The factory constructions simulate the acceptance decision directly, without estimating the unknown biases. Our  applications below use a fixed formula with a fixed
number of inputs, so the expected number of calls is an
absolute constant.

\subsection{Coin notation and the box response}
\label{sec:formula-coins}

This subsection specifies $\coin{Box}_\lambda$, the coin
implementation of the price response $\Resp$ from
\Cref{alg:resp}. We first make explicit how each input and
output coin represents its quantity.

\paragraph{Coins and their biases.}
We write $\coin C:\coin{Coin}[p]$ to mean that each call to
$\coin C$, called a \emph{flip}, returns an independent
$\mathrm{Bern}(p)$ bit.
The expression $p$ specifies the bias; its numerical value
need not be known. In pseudocode, an assignment
$\coin C\gets\coin{Coin}[f(p_1,\ldots,p_k)]$ defines a new
coin procedure using the appropriate factory and the indicated
input coins.

For a bounded quantity $b\in[-1,1]$, define
$\operatorname{enc}(b)=(b+2)/4$.
Thus $\coin{Coin}[\operatorname{enc}(b)]$ represents $b$
with a bias in $[1/4,3/4]$.
Other quantities, including load ratios, use the biases
specified explicitly below.
For the rest of this section, fix $\kappa=1/64$.
The reweighting factors $A_{ua}$ used below lie in $[\kappa,1-\kappa]$,
and all load-ratio coins will have bias $\kappa$ times the ratio.
These encodings specify the factory formulas: a bounded
quantity is replaced by $4p-2$ and a load ratio by $p/\kappa$,
where $p$ is the corresponding input bias.

\paragraph{Representing the price.}
By \eqref{eq:tau}, the box response updates the transformed price by
\[
\tau'=A\tau+B,\qquad
\tau=\atanh x,\qquad \tau'=\atanh y,
\]
with $A,B$ as defined in \Cref{alg:resp}.
Although $x,y\in(-1,1)$, their transformed values need not
have a uniform constant bound. We therefore represent
$\tau$ by the bounded pair
\begin{equation}\label{eq:bounded-price}
\xi=\frac{\tau}{\sqrt{1+\tau^2}},\qquad
\sigma=\frac1{\sqrt{1+\tau^2}},\qquad
\xi^2+\sigma^2=1.
\end{equation}
Since $\tau=\xi/\sigma$, this pair determines the price.

\paragraph{The box-response interface.}
Fix a nonisolated vertex $j$ and use the load parameters
$r,z_h,z_k$ from \eqref{eq:vertexloadparams}, suppressing
the vertex subscript.
The routine
\begin{equation}\label{eq:box-interface}
\coin{Box}_\lambda
  (\coin{\Xi},\coin S,\coin R;\coin L_h,\coin L_k)
\end{equation}
takes five coin procedures and the known step size
$\lambda\in\{1/3,1/6\}$. Their types are
\[
\begin{array}{lll}
\coin{\Xi}:\coin{Coin}[\operatorname{enc}(\xi)],&
\coin S:\coin{Coin}[\operatorname{enc}(\sigma)],&
\coin R:\coin{Coin}[\operatorname{enc}(r)],\\
\coin L_h:\coin{Coin}[\kappa z_h],&
\coin L_k:\coin{Coin}[\kappa z_k].&
\end{array}
\]
The semicolon separates the base-state coins from the
load-ratio coins. In particular, a ratio equal to $1$ is
represented by the known coin $\coin L_1:\coin{Coin}[\kappa]$.

Let $y$ be the corresponding price returned by
$\Resp_{z,v}(h)$, with $v^x=\lambda(\mathbf 1-Lk)$.
The routine returns
\[
(\coin{\Xi}',\coin S',\coin X',\coin V')
\]
with respective biases
\[
\operatorname{enc}(\xi'),\qquad
\operatorname{enc}(\sigma'),\qquad
\operatorname{enc}(y),\qquad
\operatorname{enc}(\psi(y)).
\]
Here $(\xi',\sigma')$ is the bounded representation
\eqref{eq:bounded-price} of $\atanh y$.

\paragraph{Defining the output coins.}
We first rewrite the price update $\tau'=A\tau+B$ from
\eqref{eq:tau} as an update of the bounded pair in
\eqref{eq:bounded-price}. Since $\tau=\xi/\sigma$, we have
$\tau'=(A\xi+B\sigma)/\sigma$.
Applying \eqref{eq:bounded-price} to $\tau'$ gives
\begin{equation}\label{eq:normalized-box}
u=A\xi+B\sigma,\qquad
(\xi',\sigma')=\frac{(u,\sigma)}{\sqrt{u^2+\sigma^2}}.
\end{equation}
Substituting the expressions for $A$ and $B$ from
\Cref{lem:coef} writes this update in terms of the load
parameters $r,z_h,z_k$ from \eqref{eq:vertexloadparams}:
\begin{equation}\label{eq:sample-box}
\begin{gathered}
w=1-r+rz_h,\qquad
u=\frac{\xi+\frac\lambda2\bigl(r(z_k+1)-1\bigr)\sigma}{w},
\qquad D=\sqrt{u^2+\sigma^2},\\
\xi'=u/D,\qquad \sigma'=\sigma/D.
\end{gathered}
\end{equation}
The elementary-formula factory in \Cref{lem:factory} constructs
$\coin{\Xi}',\coin S'$ with these encoding biases.
The smooth one-variable factory in \Cref{lem:smooth-factory} then constructs
$\coin X',\coin V'$ from $\coin{\Xi}'$.
In \Cref{sec:sampler-exact}, we give the two conversion
functions and verify the conditions needed to apply both
factory lemmas, including the lower bounds on denominators
and root arguments and the smoothness of the conversion
functions. This proves the following guarantee.

\begin{lemma}[Box-response coins]\label{lem:box-coins}
For every box response used by the optimizer,
$\coin{Box}_\lambda$ supplies the four coin procedures
specified above, with their exact biases.
Each output flip uses an expected constant number of
input-coin flips and expected constant additional work,
uniformly over all such responses.
\end{lemma}

\subsection{Samplers for a reweighted distribution}
\label{sec:tilted-samplers}

The routine $\coin{Tilt}$ implements one proposal reweighting.
Fix any $q\in Q^\circ$ and weights
$A_{ua}\in[\kappa,1-\kappa]$, and let
$q^+_u=\cN(q_uA_u)$.
Write $\Out_q(u)$ for a draw from $q_u$. For nonisolated $v$,
$\In_q(v)$ draws $u\in N(v)$ with probability
$q_{uv}/\ell_v(q)$.

Given these two samplers and acceptance coins
$\coin A(u,a):\coin{Coin}[A_{ua}]$, the routine returns
\begin{equation}\label{eq:tilt-interface}
(\Out_{q^+},\In_{q^+},\coin L)
\gets\coin{Tilt}(\Out_q,\In_q,\coin A),
\end{equation}
where $\coin L(v)$ has bias
$\kappa\ell_v(q^+)/\ell_v(q)$.

Incoming samples provide load-ratio coins through the identity
\begin{equation}\label{eq:sample-load-ratio}
\frac{\ell_v(q^+)}{\ell_v(q)}
=\E_{u\sim\In_q(v)}\left[\frac{q^+_{uv}}{q_{uv}}\right].
\end{equation}
Thus one call to $\In_q(v)$ followed by a coin flip with bias
$\kappa q^+_{uv}/q_{uv}$ gives the required load-ratio coin,
without evaluating either load.

\paragraph{Construction of $\coin{Tilt}$.}
First define two helper coins. The normalizer coin
$\coin Z(u)$ draws $a\leftarrow\Out_q(u)$ and returns
$\coin A(u,a)$, so its bias is $Z_u=\sum_aq_{ua}A_{ua}$.
The coordinate-ratio coin
$\coin C(u,a):\coin{Coin}[\kappa A_{ua}/Z_u]$ uses
\Cref{lem:factory} with input coins $\coin A(u,a),\coin Z(u)$.
Its bias is $\kappa q^+_{ua}/q_{ua}$.

The routine returns the following three procedures, each
using fresh draws and flips when invoked:
\begin{enumerate}[label=(\alph*)]
\item $\Out_{q^+}(u)$ draws $a\leftarrow\Out_q(u)$ and returns
$a$ if $\coin A(u,a)$ returns one; otherwise it repeats.
\item $\In_{q^+}(v)$ draws $u\leftarrow\In_q(v)$ and returns
$u$ if $\coin C(u,v)$ returns one; otherwise it repeats.
\item $\coin L(v)$ draws $u\leftarrow\In_q(v)$ and returns
the bit $\coin C(u,v)$.
\end{enumerate}
The next lemma gives the guarantee for this construction;
its proof is in \Cref{sec:sampler-exact}.

\begin{lemma}[Bounded tilt]\label{lem:tilt}
For every $q\in Q^\circ$ and weights
$A_{ua}\in[\kappa,1-\kappa]$, $\coin{Tilt}$ supplies
$\Out_{q^+},\In_{q^+},\coin L$ with the specified laws.
Each invocation of a returned procedure uses an expected
$O(1)$ calls to $\Out_q,\In_q,\coin A$ and expected
$O(1)$ additional work, uniformly over $q,A$ and the
queried vertex.
\end{lemma}

\subsection{The recursive sampling round}
\label{sec:sampler-round}

We now implement one round of the optimizer in
\Cref{alg:iteration} using $\coin{Box}_\lambda$ from
\eqref{eq:box-interface} and $\coin{Tilt}$ from
\eqref{eq:tilt-interface}.
Given the current samplers and coins, we construct a sampler
for the proposal distribution $q'_t$ produced by the gradient
stage, and the samplers and coins representing the next state.

\paragraph{What a round receives and returns.}
Write $\mathcal S_t$ for the following family of procedures:
\begin{equation}\label{eq:sampler-state}
\mathcal S_t=
(\Out_t,\In_t,\coin{\Xi}_t,\coin S_t,
 \coin X_t,\coin V_t,\coin R_t,\coin P_t).
\end{equation}
Their laws are specified below, where
$r_{t,v}=\ell_v(q_t)/(1+\ell_v(q_t))$ and
$\rho_t=\bar q_t/q_t$.
\begin{center}
\begin{tabular}{ll}
procedure & output law or coin type\\ \hline
$\Out_t(u)$ & $a\in I_u$ with probability $q_{t,ua}$\\
$\In_t(v)$ & $u\in N(v)$ with probability $q_{t,uv}/\ell_v(q_t)$\\
$\coin{\Xi}_t(v)$ & $\coin{Coin}[\operatorname{enc}(\xi_{t,v})]$\\
$\coin S_t(v)$ & $\coin{Coin}[\operatorname{enc}(\sigma_{t,v})]$\\
$\coin X_t(v)$ & $\coin{Coin}[\operatorname{enc}(x_{t,v})]$\\
$\coin V_t(v)$ & $\coin{Coin}[\operatorname{enc}(\psi(x_{t,v}))]$\\
$\coin R_t(v)$ & $\coin{Coin}[\operatorname{enc}(r_{t,v})]$\\
$\coin P_t(u,a)$ & $\coin{Coin}[\rho_{t,ua}/(2e)]$
\end{tabular}
\end{center}
One round defines
\[
\mathcal S_t\longmapsto(\Out'_t,\mathcal S_{t+1}),
\qquad \Out'_t(u)\sim q'_{t,u}.
\]
Only $\mathcal S_{t+1}$ is passed to the next round.
The other coins and samplers defined during the round serve
its internal computations. All target laws are deterministic
functions of $G$; every invocation uses fresh independent
randomness. Incoming samplers and state coins are needed
only at nonisolated vertices.

\paragraph{Generating one proposal.}
To sample from the optimizer's average proposal distribution,
we choose a round $t$ uniformly and draw from $q'_t$ using
$\Out'_t$. For $0<\delta<1$, let
\[
T_\delta=\left\lceil\frac{3B_\Delta}{\delta}\right\rceil,
\qquad q^{(\delta)}=\frac1{T_\delta}\sum_{t<T_\delta}q'_t.
\]
\begin{algorithm}[H]
\small
\DontPrintSemicolon
\caption{$\Out^{(\delta)}(u)$: one exact proposal}
\label{alg:sampler}
\KwIn{Vertex $u$, degree bound $\Delta$, and accuracy $\delta$.}
\KwOut{A neighbor or dummy with law $q^{(\delta)}_u$.}
\lIf{$\deg(u)=0$}{\KwRet{$\dmy_u$}}
$T\gets\lceil3B_\Delta/\delta\rceil$\;
Choose $t$ uniformly from $\{0,\ldots,T-1\}$\;
\KwRet{$\Out'_t(u)$}
  \tcp*{invoke the recursive procedure in \Cref{alg:sampler-round}}
\end{algorithm}

An invocation evaluates only the procedures it calls,
recursing to earlier rounds until it reaches the explicit
initial procedures below. There is no loop over all vertices
or preprocessing of the rounds.
Internal rejection trials repeat until acceptance, so the
output is one exact neighbor-or-dummy sample.
The dummy is an ordinary outcome of the proposal distribution.

\paragraph{Initialization.}
After a degree query, $\Out_0(u)$ returns a uniformly random
neighbor with probability $\deg(u)/(2\Delta^2)$, and returns
$\dmy_u$ otherwise. The sampler $\In_0(v)$ returns a uniformly
random neighbor. Since $\ell_v(q_0)=\deg(v)/(2\Delta^2)$,
the bias of $\coin R_0(v)$ is known after one degree query.
The initial transformed price is
$\tau_0=-\tfrac12\log(2\Delta-1)$.
It gives known biases for
$\coin{\Xi}_0,\coin S_0,\coin X_0,\coin V_0$, while
$\coin P_0$ has bias $1/(2e)$.
Each initial procedure uses at most two graph queries and
constant work.

\paragraph{How the four price responses are used.}
All four box calls use the interface \eqref{eq:box-interface}
with the base coins
$\coin{\Xi}_t,\coin S_t,\coin R_t$.
The first response supplies the coin for $\psi(x^*)$ used
to reweight $q_t$ into $q'_t$. The second supplies the
tentative-price coin for $x'$, used in the extragradient
scores. The third supplies the coin for $\psi(x_a)$ used
in the main proposal update. The fourth supplies the price
coins carried into $\mathcal S_{t+1}$; its coin for
$\psi(x_{t+1})$ is also used in the auxiliary update.

For clarity, the three scores from \Cref{sec:opt-sampling}
are repeated here. On a real coordinate $(u,v)$ they are
\begin{equation}\label{eq:sampling-scores}
\begin{aligned}
c^g_{uv}
 &=\frac1{16}\left[
   \frac{x_{t,v}-1}{3}
   +2\bigl(\psi(x^*_v)-\psi(x_{t,v})\bigr)\right],\\
a_{uv}
 &=\frac1{64}\left[
   \frac{x'_v-1}{6}
   +2\bigl(\psi(x_{a,v})-\psi(x_{t,v})\bigr)\right],\\
\hat s_{uv}
 &=\frac1{64}\left[
   \frac{x'_v-1}{6}
   +2\bigl(\psi(x_{t+1,v})-\psi(x_{t,v})\bigr)\right].
\end{aligned}
\end{equation}
All three scores are zero on dummy coordinates.
These formulas specify the biases of the acceptance coins;
the scores are not evaluated numerically.

\paragraph{One round.}
\Cref{alg:sampler-round} gives the recursive definitions.
A dash denotes an unused output procedure.
Beside each acceptance-coin definition we list the coins
whose biases occur in its formula.
The known coin $\coin L_1:\coin{Coin}[\kappa]$ represents
load ratio $1$. The other load ratios are
\[
z'=\frac{\ell(q'_t)}{\ell(q_t)},\qquad
\bar z=\frac{\ell(\bar q_t)}{\ell(q_t)},\qquad
z^+=\frac{\ell(q_{t+1})}{\ell(q_t)}.
\]
All definitions are pointwise at the indicated vertex or
proposal coordinate.

\begin{algorithm}[H]
\small
\DontPrintSemicolon
\caption{One sampling round: defining $(\Out'_t,\mathcal S_{t+1})$}
\label{alg:sampler-round}
\KwIn{The samplers and coins $\mathcal S_t$ in \eqref{eq:sampler-state}.}
\KwOut{A sampler $\Out'_t$ for $q'_t$, and the next family $\mathcal S_{t+1}$.}
\textbf{Gradient stage: proposals $q'_t$}\;
\Indp
$(-,-,-,\coin V^*)\gets
 \coin{Box}_{1/3}(\coin{\Xi}_t,\coin S_t,\coin R_t;
                 \coin L_1,\coin L_1)$
 \tcp*{\eqref{eq:box-interface}}
$\coin A^g(u,a)\gets\coin{Coin}[e^{-c^g_{ua}-1}]$
 \tcp*{uses $\coin X_t,\coin V_t,\coin V^*$}
$(\Out'_t,-,\coin L')\gets\coin{Tilt}(\Out_t,\In_t,\coin A^g)$
 \tcp*{\eqref{eq:tilt-interface}; $\coin L':\coin{Coin}[\kappa z']$}
$(-,-,\coin X',-)\gets
 \coin{Box}_{1/3}(\coin{\Xi}_t,\coin S_t,\coin R_t;
                 \coin L',\coin L_1)$\;
\Indm
\BlankLine
\textbf{Extragradient stage: next main state}\;
\Indp
Define $\bar{\coin L}(v)$: draw $u\leftarrow\In_t(v)$ and return the product of
 the bit returned by $\coin P_t(u,v)$ and an independent
 $\mathrm{Bern}(2e\kappa)$ bit
 \tcp*{bias $\kappa\bar z_v$}
$(-,-,-,\coin V_a)\gets
 \coin{Box}_{1/6}(\coin{\Xi}_t,\coin S_t,\coin R_t;
                 \bar{\coin L},\coin L')$\;
$\coin A^m(u,a)\gets
 \coin{Coin}[\sqrt{\rho_{t,ua}/(2e)}\,e^{-a_{ua}-1}]$
 \tcp*{uses $\coin P_t,\coin X',\coin V_a,\coin V_t$}
$(\Out_{t+1},\In_{t+1},\coin L^+)
 \gets\coin{Tilt}(\Out_t,\In_t,\coin A^m)$
 \tcp*{$\coin L^+:\coin{Coin}[\kappa z^+]$}
$(\coin{\Xi}_{t+1},\coin S_{t+1},\coin X_{t+1},\coin V_{t+1})
 \gets\coin{Box}_{1/6}(\coin{\Xi}_t,\coin S_t,\coin R_t;
                     \coin L^+,\coin L')$\;
$\coin R_{t+1}\gets
 \coin{Coin}[\operatorname{enc}(rz^+/(1-r+rz^+))]$
 \tcp*{uses $\coin R_t,\coin L^+$}
\Indm
\BlankLine
\textbf{Extragradient stage: next auxiliary ratio}\;
\Indp
$\coin A^\rho(u,a)\gets
 \coin{Coin}[(\rho_{t,ua}/(2e))^{1/4}
            e^{3a_{ua}/2-\hat s_{ua}-1}]$\;
\tcp{Uses $\coin P_t,\coin X',\coin V_a,\coin V_t,\coin V_{t+1}$.}
Define $\coin Z^\rho(u)$: draw $a\leftarrow\Out_{t+1}(u)$
 and return the bit from $\coin A^\rho(u,a)$\;
$\coin P_{t+1}(u,a)\gets
 \coin{Coin}[A^\rho_{ua}/(2eZ^\rho_u)]$
 \tcp*{uses $\coin A^\rho,\coin Z^\rho$}
\Indm
\BlankLine
\KwRet{$(\Out'_t,\mathcal S_{t+1})$}\;
\end{algorithm}

The extra known-probability flip in $\bar{\coin L}$ puts
its bias on the same scale as $\coin L'$ and $\coin L^+$.
The definitions of $\mathcal S_{t+1}$ refer to their input
procedures, which are evaluated on demand in later invocations.
No sampled bits are stored as the state.

\subsection{Exactness of the recursive sampler}
\label{sec:sampler-exact}

We first justify the two subroutine interfaces, then prove
that their composition follows the optimizer exactly.

\paragraph{Box-response coins.}
The coefficient bounds from \Cref{sec:opt-sampling} keep
the normalization in \eqref{eq:sample-box} away from zero.

\begin{lemma}[Bounded normalization]\label{lem:normalized-box}
For every box response used by the optimizer,
\begin{equation}\label{eq:D}
\tfrac1{400}\le u^2+\sigma^2\le21.
\end{equation}
\end{lemma}
\begin{proof}
By \Cref{lem:ratios,lem:coef}, $A\in[e^{-1},e]$ and $|B|<2$.
Using $\xi^2+\sigma^2=1$, Cauchy--Schwarz gives
\[
u^2+\sigma^2\le A^2+B^2+1<21.
\]
Conversely, $\xi=(u-B\sigma)/A$ gives
\[
1\le\left(\frac{1+B^2}{A^2}+1\right)(u^2+\sigma^2)
\le(5e^2+1)(u^2+\sigma^2)<400(u^2+\sigma^2).
\]
\end{proof}

\begin{proof}[Proof of \Cref{lem:box-coins}]
The normalization represents the price response exactly:
$\xi'/\sigma'=A\xi/\sigma+B=\tau'$.
By \Cref{lem:coef,lem:normalized-box}, $w\ge e^{-1}$,
the root argument is at least $1/400$, and $D\ge1/20$.
The output encoding biases lie in $[1/4,3/4]$, so
\Cref{lem:factory} constructs $\coin{\Xi}',\coin S'$.

To construct the price coins, use the fixed functions
\begin{equation}\label{eq:price-readouts}
\xi\longmapsto\tanh\!\left(\frac{\xi}{\sqrt{1-\xi^2}}\right),
\qquad
\xi\longmapsto\psi\!\left(
 \tanh\!\left(\frac{\xi}{\sqrt{1-\xi^2}}\right)\right).
\end{equation}
Applied to $\xi'$, they give $y$ and $\psi(y)$.
At $\xi=\pm1$ and beyond, extend the first function by
$\pm1$, respectively, and the second by $\log2$ on both sides.
These extensions are smooth. Indeed, writing
$\tau=\xi/\sqrt{1-\xi^2}$ gives
\[
\psi(\tanh\tau)=\tau\tanh\tau-\log\cosh\tau.
\]
Both functions approach their endpoint values exponentially
fast in $|\tau|$. Differentiating with respect to $\xi$
applies $(1+\tau^2)^{3/2}\,d/d\tau$, so every positive-order
derivative still tends to zero at the endpoints.
In particular, the extensions have four continuous derivatives
with fixed bounds. Their values and second derivatives have
fixed formulas, with the constant branches supplying the
endpoint values and derivatives.

Applying either extended function to $\xi=4p-2$ and encoding
the result gives a fixed smooth function of $p\in[0,1]$
with values in $[1/4,3/4]$.
Thus \Cref{lem:smooth-factory} constructs
$\coin X',\coin V'$ from $\coin{\Xi}'$.
Each factory has constant expected input and additional cost,
as does their composition.
\end{proof}

\paragraph{Updating proposal samplers.}
The bounded-tilt routine combines the rejection step with
the load-ratio identity \eqref{eq:sample-load-ratio}.

\begin{proof}[Proof of \Cref{lem:tilt}]
The outgoing sampler is the rejection step in
\eqref{eq:sampling-tilt}. Its acceptance probability is
$Z_u$, which is also the bias of $\coin Z(u)$.
Since $A_{ua},Z_u\in[\kappa,1-\kappa]$, the bias of
$\coin C(u,a)$ satisfies
\[
\kappa^2\le\frac{\kappa A_{ua}}{Z_u}\le1-\kappa.
\]
The denominator is at least $\kappa$, and the bias has
margin at least $\kappa^2$ from both endpoints.
Thus \Cref{lem:factory} applies with constant parameters.
An incoming trial accepts $u$ with mass
\[
\frac{q_{uv}}{\ell_v(q)}\,
\kappa\frac{q^+_{uv}}{q_{uv}}
=\frac{\kappa q^+_{uv}}{\ell_v(q)}.
\]
Its conditional law is $\In_{q^+}(v)$.
Its acceptance probability, also the bias of $\coin L(v)$, is
$\kappa\ell_v(q^+)/\ell_v(q)\ge\kappa^2$.
Both rejection loops therefore use an expected constant
number of trials, proving the exactness and cost claims.
\end{proof}

\paragraph{One recursive round.}
It remains to check the acceptance coins and the two state
coins $\coin R_{t+1},\coin P_{t+1}$.

\begin{lemma}[Exactness]\label{lem:exact}
For every $t$, \Cref{alg:sampler-round} defines
$\Out'_t$ with law $q'_t$ and a procedure family
$\mathcal S_{t+1}$ representing the next optimization state
exactly. Every intermediate coin has its stated bias.
\end{lemma}
\begin{proof}
Induct on $t$, starting from the explicit initialization
and following the dependency order in the algorithm.
All target laws are deterministic functions of $G$, and every
invocation uses fresh randomness.

By \Cref{lem:ratios}(c), the load ratios
$\bar z,z',z^+$ and the coordinate ratios lie in
$[e^{-1},e]$. The auxiliary load coin has the stated bias:
\[
\E[\bar{\coin L}(v)]
=2e\kappa\,
 \E_{u\sim\In_t(v)}\left[\frac{\rho_{t,uv}}{2e}\right]
=\kappa\frac{\ell_v(\bar q_t)}{\ell_v(q_t)}.
\]
Thus \Cref{lem:box-coins} supplies the exact price coins
for each response.

The root arguments $\rho/(2e)$ are at least $1/(2e^2)$.
The score bounds in \Cref{lem:ratios}(a) give
\begin{align*}
A^g&\in(e^{-9/8},e^{-5/6}),\\
A^m&\in[2^{-1/2}e^{-49/24},\,2^{-1/2}e^{-23/24}],\\
A^\rho&\in[2^{-1/4}e^{-77/48},\,2^{-1/4}e^{-43/48}].
\end{align*}
These intervals lie in $[\kappa,1-\kappa]$.
Consequently \Cref{lem:factory} constructs the exact weight
coins from the listed inputs, and \Cref{lem:tilt} gives the
outgoing and incoming samplers and the load-ratio coins.

For the load-state update,
$1-r+rz^+\in[e^{-1},e]$ and
$r^+=rz^+/(1-r+rz^+)\in[0,1)$.
Thus the denominator has a constant lower bound and the
encoding bias lies in $[1/4,3/4]$.
For the auxiliary update, $A^\rho$ is a constant multiple
of $W$ in \Cref{prop:reduced}, so
\[
\frac{A^\rho_{ua}}{Z^\rho_u}=\rho_{t+1,ua}\in[e^{-1},e].
\]
Here $Z^\rho_u\ge\kappa$, and the bias of $\coin P_{t+1}$
lies in $[1/(2e^2),1/2]$.
The elementary-formula factory therefore supplies both
state coins with their exact biases.
\end{proof}

\subsection{Query cost and arithmetic work}
\label{sec:sampler-cost}

Each output procedure supplied by $\coin{Box}_\lambda$ or
$\coin{Tilt}$, and each remaining formula coin, uses an
expected constant number of input calls and expected
constant additional work. These costs compound over $T$
rounds to a bound exponential in $T$, with an absolute base.
This also bounds the number of graph queries.
\Cref{prop:finite-sampling} gives the finite-arithmetic
implementation and establishes the corresponding
computational-work bound in the model of \Cref{sec:model}.

\begin{lemma}[Adaptive composition]\label{lem:compose}
Suppose a procedure makes an expected at most $k$ calls to
earlier procedures, at adaptively chosen arguments. If each
call has conditional expected cost at most $c$, given the
history before it, and the procedure's own expected cost is
at most $b$, then its expected cost is at most $b+kc$.
\end{lemma}
\begin{proof}
Let $I_j$ indicate that the $j$-th call is made and let $C_j$
be its cost, zero if the call is not made. Conditional on the
history when the call is decided, $\E C_j\le cI_j$.
The tower property and Tonelli's theorem give
$\E\sum_jC_j\le c\,\E\sum_jI_j\le ck$.
\end{proof}

\begin{proposition}[Cost of the recursive procedures]
\label{prop:cost}
There is an absolute constant $A_0\ge2$ such that every
round-$t$ procedure, and every intermediate procedure in that
round, terminates almost surely and has total work $W$
satisfying
\[
\E W\le A_0^{t+1}.
\]
The expectation is uniform over the argument and over the
history before a fresh invocation. In particular, the same
bound holds for the expected number of graph queries.
\end{proposition}
\begin{proof}
By \Cref{lem:box-coins,lem:tilt,lem:factory}, each new
procedure makes an expected constant number of input calls
and performs expected constant additional work.
The dependency order in \Cref{alg:sampler-round} has constant
depth within one round.
Applying \Cref{lem:compose} through these dependencies bounds
each round's expected cost by a fixed constant times one plus
the preceding round's maximum expected cost. Initialization
has constant cost, giving the displayed bound after choosing
$A_0$ sufficiently large. Fresh independent randomness gives
the same bounds conditional on the history before an invocation.
Finite expected total work implies almost sure termination.
\end{proof}

\begin{theorem}[Local proposal sampler]\label{thm:sampler}
For every graph $G$ and $0<\delta<1$, \Cref{alg:sampler}
returns an exact sample from $q^{(\delta)}_u$, where
\[
(2-\delta)\nu_f(G)\le P(q^{(\delta)})\le2\nu_f(G).
\]
It terminates almost surely and has expected total work at most
\[
C(\delta,\Delta):=A_0^{T_\delta+1}
=(2\Delta)^{O(1/\delta)}.
\]
The same bound holds for the expected number of graph queries.
Independent invocations use independent randomness.
\end{theorem}
\begin{proof}
Choosing a uniform round samples the average distribution
exactly by \Cref{lem:exact}. Its accepted-mass bound is
\Cref{thm:rate}; its cost follows from \Cref{prop:cost},
absorbing the constant overhead of choosing a round into
$A_0$. Finally, $T_\delta=O(\delta^{-1}\log(2\Delta))$.
\end{proof}

\section{From samples to a fractional matching}\label{sec:prop}

We use the sampler of \Cref{sec:sampling} to construct an
LCA for a fractional matching, with a worst-case work bound
per query. Each invocation of the sampler has a fixed work budget, that is exceeded, the dummy is returned.
Each vertex makes $K$ such proposals (invocations) and keeps at most
$K$ incoming ones. Giving weight $1/(2K)$ to every kept
proposal is feasible in every graph. We bound the loss from
these two steps and give consistent local access to the kept
proposals. Estimation and rounding use these guarantees
separately.

\subsection{Capping the work of a proposal}\label{sec:capping}
In the sequel we use the exact finite-arithmetic
sampler of \Cref{prop:finite-sampling}.
Increasing the absolute constant $A_0$ in
$C(\delta,\Delta)$ if necessary, this quantity also bounds
its expected computational work in the model of
\Cref{sec:model}.
All work budgets below count this computational work.

The expected total-work bound of \Cref{thm:sampler} gives a
worst-case work budget directly. For $0<\delta,\tau<1$, set
\begin{equation}\label{eq:work-cap}
B=\left\lceil\frac{C(\delta,\Delta)}{\tau}\right\rceil.
\end{equation}
The \emph{capped sampler} runs $\Out^{(\delta)}(u)$ for at
most $B$ units of work. If the sampler has returned by then,
we use its output; otherwise we abort and return $\dmy_u$.
All recursive calls share this one budget, including the
work inside Bernoulli factories. Maintaining the work
counter adds only constant overhead per operation, so the
total work is $O(B)$. Write $\tilde q_u$ for the capped
output law.

\begin{lemma}[Capping]\label{lem:cap}
The capped sampler has a deterministic $O(B)$ total-work
bound and satisfies
\[
(2-\delta-2\Delta\tau)\nu_f(G)\le P(\tilde q)\le2\nu_f(G).
\]
\end{lemma}
\begin{proof}
Couple the capped and uncapped runs using the same randomness.
If $W$ is the uncapped run's total work, then
\Cref{thm:sampler} and Markov's inequality give
\[
\Prob\{\text{abort}\}=\Prob\{W>B\}
\le\frac{\E W}{B}\le\tau.
\]
Capping only replaces outcomes by the dummy. Thus, for every
$u$, the real-coordinate probabilities can only decrease,
and their total decrease is at most $\tau$:
\[
\tilde q_{uv}\le q^{(\delta)}_{uv}\quad(v\in N(u)),
\qquad
\sum_{v\in N(u)}
\bigl(q^{(\delta)}_{uv}-\tilde q_{uv}\bigr)\le\tau.
\]
Since $\ell\mapsto\min\{1,\ell\}$ is monotone and
$1$-Lipschitz,
\[
0\le P(q^{(\delta)})-P(\tilde q)
\le\sum_v\bigl(\ell_v(q^{(\delta)})-\ell_v(\tilde q)\bigr)
\le\tau n_0\le2\Delta\tau\nu_f(G).
\]
Here isolated vertices send no real mass, and the last
inequality is \Cref{lem:active}. The lower bound follows
from \Cref{thm:sampler}; the upper bound follows from
\Cref{lem:value}(a).
\end{proof}

For the rest of the construction, choose
\begin{equation}\label{eq:proposal-parameters}
\delta=\frac\eps{32},\qquad \tau=\frac\eps{64\Delta},
\qquad K=2^{\lceil\log_2(4096/\eps^2)\rceil}.
\end{equation}
Then $K=\Theta(\eps^{-2})$ is a power of two, as needed
by the later degree-halving procedure. The choice of
$\tau$ bounds the loss in accepted mass from aborted
draws by $2\Delta\tau\nu_f(G)=\eps\nu_f(G)/32$, and
the work budget satisfies $B=(2\Delta)^{O(1/\eps)}$.

\subsection{Keeping proposals}\label{sec:kept-proposals}

Each vertex $u$ makes $K$ independent capped proposals,
identified by $(u,j)$ for $j=1,\ldots,K$. A real proposal to
$v$ is a copy of the edge $uv$ with this ID. Let $N_v$ count
the proposals received by $v$. The receiver keeps the
$\min\{K,N_v\}$ proposals with smallest IDs, in lexicographic
order. All kept copies form a multigraph $H$ on $V$, with
\[
S:=|E(H)|=\sum_v\min\{K,N_v\}.
\]
Parallel copies, including proposals in opposite directions,
remain distinct. A vertex has at most $K$ kept outgoing and
$K$ kept incoming copies, so $H$ has maximum degree at most
$2K$. Thus the weights
\begin{equation}\label{eq:proposal-fractional}
f^H_{\{u,v\}}
=\frac{\#\{\text{kept copies of }uv\}}{2K}
\end{equation}
form a fractional matching of $G$, of value $S/(2K)$.
The multigraph $H$ is bipartite whenever $G$ is bipartite.

Clipping the incoming count at $K$ loses little in expectation.
The following estimate isolates that loss.

\begin{lemma}[Relative clipping]\label{lem:clip}
If $N$ is a sum of independent Bernoulli variables with mean
$K\ell$, for an integer $K\ge1$, then
\[
\left(1-\frac2{\sqrt K}\right)\min\{1,\ell\}
\le\frac{\E\min\{K,N\}}K
\le\min\{1,\ell\}.
\]
\end{lemma}
\begin{proof}
The upper bound is Jensen's inequality. Suppose first that
$\mu:=\E N\le K$. Write $N=\sum_iX_i$, $p_i=\E X_i$,
and $N_i=N-X_i$. Since $(N-K)_+$ vanishes unless $N>K$,
\begin{align*}
\E(N-K)_+
&\le\frac1K\E[N(N-K)_+]\\
&=\frac1K\sum_i p_i\E(N_i+1-K)_+\\
&\le\frac\mu K(\sqrt K+1)
\le\frac{2\mu}{\sqrt K}.
\end{align*}
For the penultimate step, $\E N_i\le K$ and
$\operatorname{Var}N_i\le K$ imply
$\E(N_i+1-K)_+\le\E|N_i-\E N_i|+1\le\sqrt K+1$.
Subtracting the loss from $\mu$ proves this case.
If $\mu>K$, independently retain each success with probability
$K/\mu$. The thinned count has mean $K$ and is at most $N$,
so the first case gives the required bound.
\end{proof}

\begin{corollary}[Expected value]\label{cor:ES}
The fractional matching of kept proposals satisfies
\[
(1-\eps/16)\nu_f(G)
\le\frac{\E S}{2K}\le\nu_f(G).
\]
\end{corollary}
\begin{proof}
Each $N_v$ is a sum of $K\deg(v)$ independent indicators
with mean $K\ell_v(\tilde q)$. Summing \Cref{lem:clip}
and applying \Cref{lem:cap} gives
\[
\frac{\E S}{2K}
\ge\left(1-\frac2{\sqrt K}\right)\frac{P(\tilde q)}2
\ge\left(1-\frac2{\sqrt K}\right)
     \left(1-\frac\delta2-\Delta\tau\right)\nu_f(G).
\]
By \eqref{eq:proposal-parameters},
$2/\sqrt K\le\eps/32$ and
$\delta/2+\Delta\tau=\eps/32$, giving the lower bound.
The upper bound follows from feasibility for every realization.
\end{proof}

\subsection{Consistent local access}\label{sec:proposal-access}

Independent seeds fix the proposals, so any query can
reconstruct the same kept copies. Give each proposal $(u,j)$
its own seed $\omega_{u,j}$ as in \Cref{sec:model}.
Re-evaluating it with the same seed and work budget reproduces
the same outcome. These seeds therefore fix one multigraph
$H$, regardless of the query order.

\Cref{alg:proposal-access} gives the local routines.
\texttt{Incoming}$(v)$ lists the copies kept by $v$;
\texttt{List}$(v)$ also checks which proposals made by $v$
were kept by their receivers. Each list stores the IDs and
endpoints of its copies. The routines recompute the entries
they need and use no persistent memory.

\begin{algorithm}[H]
\small
\DontPrintSemicolon
\SetKwFunction{Propose}{Propose}
\SetKwFunction{Incoming}{Incoming}
\SetKwFunction{ListH}{List}
\SetKwProg{Fn}{Procedure}{:}{}
\caption{Local access to the proposal multigraph}
\label{alg:proposal-access}
\KwIn{Fixed parameters $\delta,B,K$ and independent
proposal seeds $\omega_{u,j}$.}

\Fn{\Propose{$u,j$}}{
  Run $\Out^{(\delta)}(u)$ with seed $\omega_{u,j}$
  for at most $B$ units of total sampler work\;
  \lIf{the sampler has returned an output $a$}{\KwRet{$a$}}
  \KwRet{$\dmy_u$}\;
}
\BlankLine
\Fn{\Incoming{$v$}}{
  $L\gets$ empty list\;
  \ForEach{$u\in N(v)$}{
    \For{$j=1,\ldots,K$}{
      \If{\Propose{$u,j$}$=v$}{
        Append the copy of $uv$ with ID $(u,j)$ to $L$\;
      }
    }
  }
  Sort $L$ by ID and return its first $\min\{K,|L|\}$ copies\;
}
\BlankLine
\Fn{\ListH{$v$}}{
  $L\gets$ \Incoming{$v$}\;
  \For{$j=1,\ldots,K$}{
    $w\gets$ \Propose{$v,j$}\;
    \If{$w\ne\dmy_v$}{
      \If{the copy with ID $(v,j)$ belongs to \Incoming{$w$}}{
        Append that copy to $L$\;
      }
    }
  }
  Sort $L$ by ID and return $L$\;
}
\end{algorithm}

\begin{lemma}[The proposal multigraph]\label{lem:H}
For every graph $G$, $H$ has $S$ edges and maximum degree at
most $2K$. Giving each copy weight $1/(2K)$ yields a feasible
fractional matching with the expected value in \Cref{cor:ES}.
The list of copies incident to a specified vertex, sorted by
ID, can be computed consistently using at most
\[
A=O\!\left(\Delta K^2\bigl(B+\log(2\Delta K)\bigr)\right)
=(2\Delta)^{O(1/\eps)}
\]
total work, and hence at most that many graph queries.
\end{lemma}
\begin{proof}
The size, degree, and feasibility claims follow from the
construction, and \Cref{cor:ES} gives the expected value.

A call to \texttt{Incoming} evaluates at most $\Delta K$
proposals and sorts at most that many copies. Its total work
is $O(\Delta K(B+\log(2\Delta K)))$.
A call to \texttt{List} makes at most $K+1$ such calls,
evaluates another $K$ proposals, and sorts at most $2K$
kept copies. This gives the stated bound.
Every decision uses the same IDs and proposal seeds, so all
evaluations agree. Since $G$ has no loops, an incoming copy
and an outgoing copy at a vertex are distinct, and
\texttt{List} returns exactly its incident copies in $H$.
\end{proof}

\section{The size estimator}\label{sec:est}

The estimator samples the incoming contribution of a uniformly
random vertex. Recall that the fractional matching of
\Cref{sec:prop} has value
\[
\frac{S}{2K}=\frac12\sum_v\frac{\min\{K,N_v\}}K.
\]
One \emph{estimation trial} chooses a uniform vertex $v$,
generates its incoming proposals, and returns the normalized
count $\min\{K,N_v\}/K$. Averaging independent trials and
multiplying by $n/2$ estimates the expected value of this
fractional matching, which is close to $\nu_f(G)$ by
\Cref{cor:ES}.

Write $\coin{CappedOut}_{\delta,B}(u)$ for the capped sampler
of \Cref{sec:capping}. Each invocation uses fresh independent
randomness, even when a source or receiver is sampled again.
The estimator only counts incoming proposals; it needs no
proposal IDs, sorting, shared seeds, or rounding.

\begin{algorithm}[H]
\small
\DontPrintSemicolon
\SetAlgoVlined
\SetAlgoNoEnd
\SetKwFunction{CappedOut}{CappedOut}
\caption{Estimating the fractional matching number}
\label{alg:estimator}
\KwIn{Graph access, $n$, degree bound $\Delta$, and
$0<\eps,\zeta<1/2$.}
\KwOut{An estimate $\widehat\nu_f$ of $\nu_f(G)$.}

Set $\delta,\tau,K$ as in \eqref{eq:proposal-parameters}\;
$B\gets\lceil C(\delta,\Delta)/\tau\rceil$\;
$s\gets\lceil2\eps^{-2}\log(2/\zeta)\rceil$\;
\BlankLine
\For{$i=1,\ldots,s$}{
  Choose $v$ uniformly from $V$\;
  $N\gets0$\;
  \ForEach{$u\in N(v)$}{
    \For{$j=1,\ldots,K$}{
      $a\gets$ \CappedOut{$u;\delta,B$}
        \tcp*{fresh randomness}
      \lIf{$a=v$}{$N\gets N+1$}
    }
  }
  $Z_i\gets\min\{K,N\}/K$\;
}
\KwRet{$\displaystyle\widehat\nu_f=
  \frac{n}{2s}\sum_{i=1}^s Z_i$}\;
\end{algorithm}

On bipartite graphs, the same output estimates $\nu(G)$.

\begin{proof}[Proof of \Cref{thm:randomized}(b)]
For a fixed $v$, the trial has exactly the incoming-count law
used in \Cref{sec:prop}. Thus \Cref{cor:ES} gives
\[
\frac n2\E Z\in[(1-\eps/16)\nu_f(G),\nu_f(G)]
\subseteq[\nu_f(G)-\eps n/32,\nu_f(G)].
\]
The inclusion uses $\nu_f(G)\le n/2$. Because the trials
are independent and lie in $[0,1]$,
\Cref{thm:mcd} yields
\[
\Prob\left\{\left|\frac1s\sum_iZ_i-\E Z\right|
                         >\eps/2\right\}
\le2e^{-s\eps^2/2}\le\zeta.
\]
Otherwise the sampling error is at most $\eps n/4$, and the
total error is at most $9\eps n/32$, in particular at most
$\eps n$. Each trial uses $O(\Delta K(B+1))$ total work.
The resulting worst-case bound on both graph queries and
arithmetic work is
\[
O\!\left(\Delta K(B+1)\eps^{-2}\log(2/\zeta)\right)
=(2\Delta)^{O(1/\eps)}\log(2/\zeta).
\]
\end{proof}

\begin{remark}[Estimation from the deterministic oracle]
\label{rem:est-from-lca}
The estimation claim in \Cref{thm:deterministic} follows by
the same vertex-sampling argument. Use
\Cref{cor:fractional-lca} with relative error $\eps/2$, and for a
uniform vertex $v$ compute its incident weight
$Y_v=\sum_{e\ni v}f_e\in[0,1]$.
Then $(n/2)\E Y_v=\sum_e f_e$, whose deficit from
$\nu_f(G)$ is at most $\eps\nu_f(G)/2\le\eps n/4$.
Averaging $O(\eps^{-2}\log(2/\zeta))$ independent vertex
samples estimates this value within $\eps n/2$ with
probability at least $1-\zeta$. Evaluating $Y_v$ makes at
most $\Delta$ edge-weight queries; these polynomial factors
are absorbed into the stated bound. Only vertex sampling is
randomized.
\end{remark}

\section{Rounding to a consistent local oracle}\label{sec:lca}

The proposal multigraph $H$ has $S$ copies, degree at most
$D=2K$, and represents the fractional matching $f^H$ of
\Cref{sec:prop}. We round its bipartite double cover $B(H)$
to a matching and project back to $G$. The resulting output
is half-integral on every graph and integral whenever $G$ is
bipartite. The algorithm requires neither a bipartition nor
knowledge of whether $G$ is bipartite.

The key is to use the same random marks for the two lifts of
each proposal copy. On bipartite inputs, this makes the
rounding decisions agree in the two copies of each connected
component. The optimization, sampling, and proposal
construction all run directly on $G$; only the rounding uses
the double cover.

\subsection{Local degree halving}\label{sec:halving}

The rounding repeatedly halves a degree bound, doubling the
weight represented by each remaining edge. Pairing incident
copies reduces each step to choosing alternating edges along
paths and even cycles. Random marks cut these components into
pieces that can be explored with bounded work.

Set $H_0=B(H)$, so $H_0$ has $2S$ copies and maximum degree
at most $D=2K=2^J$. Each lifted copy has an ID consisting of
its original proposal ID followed by a side index; IDs are
ordered lexicographically and retained throughout the rounding.
Write $D_i=D/2^i$ and $m_i=|E(H_i)|$. Fix a loss allowance
$0<\xi<1/2$ and set
\begin{equation}\label{eq:halving-parameters}
s=\left\lceil\log_2\frac{2J}{\xi}\right\rceil,\qquad
p=2^{-s},\qquad
R=\frac{2s}{p}.
\end{equation}
This choice gives $\xi/(4J)<p\le\xi/(2J)$ and
$R\ge2\log(1/p)/p$.
Each mark is generated from exactly $s$ independent
random bits: it is one precisely when all these bits
are zero.

\paragraph{One halving step.}
At each vertex of $H_i$, sort incident copies by ID and pair
consecutive ones, leaving one unpaired if the degree is odd.
The \emph{link multigraph} $P_i$ has the copies of $H_i$ as
vertices and these pairs as edges. A copy has at most one
link at each endpoint, so $P_i$ has maximum degree two.
Along a cycle, links alternate between the two sides of
$H_0$; every cycle is therefore even.

At level $i$, give each original proposal copy an independent
$\mathrm{Bern}(p)$ mark, shared by its surviving lifts.
Marks are independent across levels and independent of the
proposal seeds. Write $B_R(e)$ for the vertices of $P_i$
within distance $R$ of a lifted copy $e$. Delete $e$ if it is
marked, or if $|B_R(e)|\ge R$ and $B_R(e)$ contains no mark.
On each surviving component, keep the larger class of its
proper $2$-coloring; on a tie, keep the class containing the
smallest ID. These kept copies form $H_{i+1}$.

\begin{lemma}[{One halving step; cf.\ \citep{Fis17,AKO18}}]
\label{lem:halve}
Every surviving component of $P_i$ has at most $2R$ vertices,
and $H_{i+1}$ has maximum degree at most $D_{i+1}$.
If $Z_i$ copies are deleted, then
$m_{i+1}\ge(m_i-Z_i)/2$.
\end{lemma}
\begin{proof}
A surviving path or cycle with more than $2R$ vertices has a
vertex whose radius-$R$ ball lies within that component.
The ball has at least $R$ vertices and no mark, so the
deletion rule would have removed this vertex.

The two copies of a pair cannot both be kept: if both
survive, they are adjacent in $P_i$ and have opposite colors.
Hence each vertex retains at most
$\lceil\deg_{H_i}(v)/2\rceil\le D_i/2=D_{i+1}$ copies,
since $D_i$ is even for $i<J$.
Choosing the larger color class keeps at least half of all
surviving copies, which proves the edge-count bound.
\end{proof}

Sharing marks preserves the expected-loss bound: any set of
$R$ lifted copies contains at least $R/2$ distinct original
IDs, and hence at least $R/2$ independent marks.

\begin{lemma}[Deletion loss]\label{lem:deleted}
Conditional on $H_i$, $\E[Z_i\mid H_i]\le2pm_i$.
\end{lemma}
\begin{proof}
A copy is marked with probability $p$. If its radius-$R$
ball has at least $R$ vertices, that ball has no mark with
probability at most
\[
(1-p)^{R/2}\le e^{-pR/2}\le p.
\]
Thus each copy is deleted with probability at most $2p$;
summing proves the claim.
\end{proof}

\begin{proposition}[{Local degree halving; cf.\ \citep{Fis17,AKO18}}]
\label{prop:halving}
After $J$ steps, $H_J$ is a matching and
\[
\E[|E(H_J)|\mid H_0]
\ge(1-\xi)\frac{|E(H_0)|}{D}.
\]
If a sorted list of copies at a vertex of $H_0$ costs at
most $A$ total work, a mate query to $H_J$ costs at most
$A(CDR^2)^J$ total work, for an absolute constant $C$.
The same bound therefore holds for graph queries and
arithmetic operations. Fixed IDs and marks indexed by
original proposal ID and level give consistent answers in
every query order.
\end{proposition}
\begin{proof}
The final degree bound is $D_J=1$, so $H_J$ is a matching.
By \Cref{lem:halve,lem:deleted},
\[
\E\!\left[\frac{m_{i+1}}{D_{i+1}}\,\middle|\,H_i\right]
\ge\frac{m_i-\E[Z_i\mid H_i]}{D_i}
\ge(1-2p)\frac{m_i}{D_i}.
\]
Iterating expectations gives
$\E[m_J\mid H_0]\ge(1-2p)^Jm_0/D\ge(1-2pJ)m_0/D$.
Since $2pJ\le\xi$, this is the value bound.

For local evaluation, let $\mathrm{List}_i(v)$ return the
incident copies in $H_i$ in ID order, at worst-case cost $A_i$.
The two possible neighbors of a copy in $P_i$ are obtained
from the lists at its endpoints. Testing deletion uses
$O(R)$ such list calls. Determining whether a surviving copy
is kept explores its component of at most $2R$ copies and
tests their boundary, using $O(R^2)$ list calls in total.
This reveals both color classes and their smallest IDs.
Generate each encountered copy's mark once and reuse it
during this exploration. The tested balls together contain
$O(R)$ copies, so generating their marks costs
$O(Rs)\le O(R^2)$ additional work.
Filtering $\mathrm{List}_i(v)$ through these tests preserves
ID order and gives
\[
A_{i+1}\le CDR^2 A_i.
\]
This bound includes comparisons and all other local work.
Finally, the fixed seeds define one global sequence
$H_0,\ldots,H_J$; every evaluation reconstructs its relevant
part using the same choices.
\end{proof}

\subsection{Integral and half-integral outputs}\label{sec:round-general}

Apply \Cref{prop:halving} with $\xi=\eps/32$, and let
$M=E(H_J)$. The incident lists of $H_0=B(H)$ are obtained
from those of $H$ in \Cref{lem:H} with constant-factor
overhead. We now project $M$ to one fractional matching of
$G$ and show that it is integral on bipartite inputs.

\paragraph{Projection and value.}
As in \Cref{lem:double}(b), each matched lift contributes
weight $1/2$ to its original edge of $G$, with contributions
from parallel copies added. At each vertex, its two lifts
contribute at most $1/2$ each. Thus the resulting weights
$f_e$ are feasible, belong to $\{0,1/2,1\}$, and satisfy
$\sum_e f_e=|M|/2$. Conditional on $H$,
\[
\E\!\left[\sum_e f_e\,\middle|\,H\right]
\ge(1-\eps/32)\frac{S}{2K}.
\]
Together with \Cref{cor:ES}, this gives
\[
\E\sum_e f_e
\ge(1-\eps/32)(1-\eps/16)\nu_f(G)
\ge(1-3\eps/32)\nu_f(G).
\]

\paragraph{Integrality on bipartite inputs.}
If $G$ is bipartite, so is $H$, and the double cover of each
connected component of $H$ consists of two isomorphic copies.
Swapping the lifts gives an isomorphism between these copies
that preserves original proposal IDs. Within either copy,
these IDs are distinct, so the secondary side indices never
affect their relative order.

The two copies therefore undergo identical rounding
decisions. Indeed, this holds initially and is preserved
at every level: incident pairings correspond, shared marks
give corresponding deletions, and the larger-class rule and
its smallest-ID tie-break choose corresponding color classes.
Consequently, $M$ contains both lifts of each proposal copy
or neither. Its projection is integral and hence is a
matching of $G$. This conclusion follows from the input
being bipartite; the algorithm never needs to test it.

\paragraph{Local queries.}
An edge query for $uv$ uses two mate queries to $M$, at
$(u,1)$ and $(v,1)$, and returns half the number of these
that are matched to $(v,2)$ and $(u,2)$, respectively.
On bipartite inputs, a mate query at $v$ uses just the mate
of $(v,1)$, returning its original vertex or reporting that
$v$ is unmatched. Both query interfaces access the same
projected output.

\subsection{Completing the LCA guarantee}\label{sec:lca-proof}

We finish \Cref{thm:randomized}(a) using the expected-value
bound above and then check total work.
Write $Y=\sum_e f_e$ and $\nu=\nu_f(G)$. Feasibility gives
$0\le Y\le\nu$, and
\[
\E(\nu-Y)\le\frac{3\eps}{32}\nu.
\]
If $\nu=0$, the output is already optimal. Otherwise,
Markov's inequality applied to the nonnegative deficit gives
\[
\Prob\{Y<(1-\eps)\nu\}
\le\frac{\E(\nu-Y)}{\eps\nu}
\le\frac3{32}<0.1.
\]
Thus the output has the required multiplicative guarantee
with probability at least $0.9$. On bipartite inputs it is
a matching and $\nu_f(G)=\nu(G)$.
Any other fixed success probability below one is obtained
by a constant rescaling of the accuracy parameters.

\paragraph{Work per query.}
\Cref{lem:H,prop:halving} bound total work
by a constant multiple of
\[
\Delta K^2\bigl(B+\log(2\Delta K)\bigr)(CDR^2)^J.
\]
Here $K,D,R$ are polynomial in $1/\eps$,
$J=O(\log(2/\eps))$, and
$(CDR^2)^J=e^{O(\log^2(2/\eps))}$.
Together with $B=(2\Delta)^{O(1/\eps)}$, this gives
$(2\Delta)^{O(1/\eps)}$ total work per output query.
The two mate queries for an edge weight add only a constant
factor. The bound is independent of $n$.

\paragraph{Validity and consistency.}
Proposal seeds fix $H$, and the marks indexed by original
proposal ID and level fix every rounding decision. Each
query therefore reconstructs part of the same projected
fractional matching, in every query order, including adaptive
ones. The proposal work cap and the bounded rounding
explorations ensure that every query terminates within the
stated budget. Feasibility and consistency hold for every
realization of the seeds; only the approximation guarantee
is probabilistic. This proves \Cref{thm:randomized}(a).

\begin{remark}[Constants]
The absolute constants in the exponent are large and are not
optimized. The cap multiplies the expected sampler cost by
$1/\tau=O(\Delta/\eps)$, and the degree-halving overhead is
$\exp(O(\log^2(2/\eps)))$; both are absorbed into
$(2\Delta)^{O(1/\eps)}$. The result concerns the dependence
of the exponent on $1/\eps$.
\end{remark}

\section{Conclusion and open questions}\label{sec:conclusion}

Our approach combines a rapidly converging local optimizer
with exact recursive sampling of its iterates. The optimizer
gives deterministic local access to a fractional matching;
the sampler reduces the dependence on the degree by replacing
neighborhood expansion with constant expected branching.
This combination suggests several directions for further work.

\paragraph{Integral matchings in general graphs.}
Can a randomized mate-query LCA find a
$(1-\eps)$-approximate matching in a general graph using
$(2\Delta)^{O(1/\eps)}$ queries and arithmetic operations
per query, with constant success probability?
Our fractional relaxation leaves an obstruction from odd
cycles, since $\nu_f(G)$ can exceed $\nu(G)$.
The recent preprint of \citet{OpenAI26} on almost-linear-time
exact matching combines representations of odd components
with continuous optimization and flow-based repairs.
Whether its treatment of these components can be adapted
to an efficiently locally simulatable method is an
interesting question. 

\paragraph{Graphs of unbounded degree.}
When the bounded-degree assumption is dropped, Bhattacharya, Kiss, and Saranurak~\cite{BKS23b} estimate $\nu(G)$ within $\pm\eps n$ in
$n^{2-\Omega_\eps(1)}$ queries, and Behnezhad, Roghani, and
Rubinstein~\cite{brr24} prove an $\Omega(n^{2-\delta(\eps)})$ lower bound. However, the dependence on $\eps$ in the exponent remains wide open. Because \cite{BKS23b} simulates McGregor's augmenting-path streaming algorithm \cite{mcgregor2005finding}, its exponent savings $\Omega_\eps(1)$ is doubly exponentially small in $1/\eps$, whereas the lower bound of~\cite{brr24} only forces $\delta(\eps)$ to vanish roughly as $1/\sqrt{\log(1/\eps)}$. Can continuous optimization methods avoid this compounding of augmenting-path phases in the unbounded-degree regime as well, and achieve $n^{2-\operatorname{poly}(\eps)}$ queries?

\paragraph{Other local optimization problems.}
It would be useful to identify a broader class of
optimization methods whose iterates admit exact recursive
sampling with constant expected branching.
Capacitated allocation, fractional packing, and
entropy-regularized transport are natural candidates.
The challenge is to obtain both fast convergence and
a representation by samplers and coins that is preserved
by the updates, with uniformly bounded simulation cost.
Our construction shows how these requirements can guide
the choice of formulation, regularizer, and update rules.

\section*{Acknowledgments}

Edith Cohen was partially supported by the Israel Science
Foundation (grant No.~1156/23).

\paragraph{AI methodology.}
We acknowledge substantive contributions from Astra (OpenAI),
Claude (Anthropic), and Gemini (Google). Starting from an
earlier unpublished writeup by the second author, these
systems contributed the first proof of an additive
approximation guarantee. Through many subsequent iterations
with the authors, they helped strengthen the guarantees,
refine the statements and presentation, and identify
connections to related work. This assistance materially
influenced the proofs, result statements, exposition,
and related-work discussion.
The authors take full responsibility for the correctness,
originality, and integrity of all content, including
the proofs and references.

\bibliographystyle{plainnat}
\bibliography{refs}

\clearpage
\appendix
\section{Bernoulli-factory proofs}\label{sec:factory-proof}

This appendix proves the two primitives stated together in
\Cref{sec:acceptance-primitive}, including their arithmetic-work
bounds. The extensions used for the price readouts are specified
in \Cref{sec:sampler-exact}; they are part of the sampler's
preparation, not an additional assumption of these proofs.

\subsection{Elementary formulas}

We represent each intermediate value by a bounded unbiased
random variable, and convert the final representation into a
coin. Every sample and every repeated trial below uses fresh
independent randomness.

\begin{proof}[Proof of \Cref{lem:factory}]
\emph{Linear amplification.}
For fixed $C>1$ and $\eta>0$, the linear factory of
\citet{Huber16} converts a $p$-coin into a $Cp$-coin whenever
$Cp\le1-\eta$, using $O(C/\eta)$ expected input flips.
Its additional work is $O(N+1)$ on an execution using $N$
input flips in our model: each inner step uses one input flip,
one known-parameter geometric draw, and constant arithmetic;
each outer step contains an inner step and adds constant work.
A geometric draw uses a uniform draw, logarithms, and integer
rounding. Thus fixed $C,\eta$ give constant expected additional
work. For $0\le C\le1$, thinning suffices.

\emph{Bounded unbiased representations.}
A procedure represents $a$ if it returns independent samples
$X$ with $\E X=a$ and $|X|\le B$ for a known constant $B>0$.
If $a\ge m>0$, draw $X$ and then a known-bias coin with
conditional bias $(B-X)/(2B)$. This is a coin of bias
\[
q=\frac{B-a}{2B}.
\]
Linear amplification with multiplier $2$ applies because
$2q=1-a/B\le1-m/B$. Complementing its output gives an
$a/B$-coin. One may use slack $\min\{1/2,m/B\}$.
This conversion uses constant expected calls and additional
work for fixed $B,m$.

\emph{Arithmetic.}
An input bit represents its bias. Constants, sums, differences,
and scalar multiples preserve bounded unbiased representations.
If $X,Y$ independently represent $a,b$, their product represents
$ab$ and is bounded by $B_XB_Y$. Every repeated occurrence in
a formula uses an independent representation; in particular,
a square uses two independent copies. The formula is fixed,
so these operations preserve constant expected cost.

\emph{Reciprocals.}
Suppose $X$ represents $a\ge m>0$, with $|X|\le B$.
Set $c=m/2$ and $D=B+c$. The variable $X-c$ has mean
$a-c\ge c$ and absolute value at most $D$, so the conversion
above supplies an $(a-c)/D$-coin.
In each trial, choose one of two branches uniformly. On the
first branch, flip a known $c/D$-coin and return $1$ on heads.
On the second, flip an $(a-c)/D$-coin and return $0$ on heads.
On tails, repeat. A trial accepts with probability
$a/(2D)\ge m/(2D)$, and the returned bit $J$ satisfies
\[
\Prob\{J=1\}=\frac{c/(2D)}{a/(2D)}=\frac ca.
\]
Thus $J/c$ represents $1/a$, is bounded by $1/c$, and has
constant expected cost. A quotient is the product of
independent numerator and reciprocal representations.

\emph{Roots.}
For $0<\alpha<1$, \citet{Mendo19} gives the following
$p^\alpha$-coin. At iteration $i=1,2,\ldots$, flip a
$p$-coin and return $1$ on heads; otherwise return $0$ with
probability $\alpha/i$, and continue if neither event occurs.
The expected number of input flips is $p^{\alpha-1}$, and
each iteration has constant additional work. If $X$ represents
$a\ge m>0$ with $|X|\le B$, convert it to an $a/B$-coin,
apply this procedure, and multiply the returned bit by
$B^\alpha$. The result represents $a^\alpha$, is bounded by
$B^\alpha$, and has constant expected cost since $a/B\ge m/B$.
We use $\alpha=1/2$ and $1/4$.

\emph{Exponentials.}
If $X$ represents $a$ with $|X|\le B$, drawing $X$ and then
a known-bias coin with conditional bias $(X+B)/(2B)$ gives
a coin of bias $p=(a+B)/(2B)$. Draw $N\sim\mathrm{Poisson}(2B)$
and let $J=1$ if all $N$ independent $p$-coins are heads,
including when $N=0$. Then
\[
\E J=\E[p^N]=e^{2B(p-1)}=e^{a-B}.
\]
Hence $e^B J$ represents $e^a$ and is bounded by $e^B$.
It uses at most $2B$ expected calls to the $X$-procedure.
The Poisson count can be sampled using exponential
interarrival times in expected $O(B+1)$ work, which is
constant for fixed $B$.

\emph{The final coin.}
Apply these constructions along the fixed formula $\Phi$.
Every intermediate has a known finite support bound. The
promised positive lower bounds permit all reciprocal and
root steps, and \Cref{lem:compose} gives constant expected
input calls and additional work for each representation.
Write $f=\Phi(p)$ and choose a support bound $B\ge2$ for
its representation. Since $f\ge\gamma$, positive-mean
conversion gives an $f/B$-coin. A final linear amplification
with multiplier $B$ and slack $\gamma$ produces an exact
$f$-coin, because $f\le1-\gamma$. All constants depend only
on the formula and its promised bounds. Finite expected
work implies almost sure termination.
\end{proof}

\subsection{Smooth one-variable functions}

The proof uses a mixture of Bernstein polynomials. Choosing
the polynomial first makes both the input-flip count and the
coefficient-evaluation work easy to bound.

\begin{proof}[Proof of \Cref{lem:smooth-factory}]
Put $g(z)=z(1-z)f''(z)$ and $b_n(z)=f(z)-g(z)/(2n)$.
Define the degree-$n$ Bernstein polynomial
\[
P_n(p)=\sum_{j=0}^n b_n(j/n)\binom nj p^j(1-p)^{n-j}.
\]
It converges uniformly to $f$. We first bound the change
between degrees $n$ and $2n$ at the coefficient level.

When $P_n$ is expressed in degree $2n$, its coefficient at
index $j$ is $\E b_n(Z)$, where $Z=H/n$ and $H$ is
hypergeometric: sample $n$ positions from $2n$ positions,
of which $j$ are marked. Writing $z=j/(2n)$, we have
\[
\E(Z-z)=\E(Z-z)^3=0,\qquad
\E(Z-z)^2=\frac{z(1-z)}{2n-1},\qquad
\E(Z-z)^4=O(n^{-2}).
\]
The odd moments vanish because a sample and its complement
have the same law. The fourth-moment bound is uniform in
$j$, as follows from the fourth central moment of a
hypergeometric count. Taylor expansion through order three
for $f$ and through order one for $g$ therefore gives
\[
\E b_n(Z)=b_{2n}(z)+O(n^{-2}),
\]
uniformly in $j$. The constant is controlled by the bounds
on the first four derivatives of $f$. Choose a fixed known
$L>0$ such that every degree-$2n$ coefficient of $P_{2n}-P_n$
has absolute value at most $L/n^2$.

Choose a sufficiently large fixed integer $n_0$, and put
\[
n_j=2^j n_0,\qquad w_j=2L/n_j^2,\qquad
W=\sum_{j\ge0}w_j.
\]
We can ensure $W\le\gamma/2$ and that all coefficients of
$P_{n_0}$ lie in $[3\gamma/4,1-3\gamma/4]$. Consequently,
\[
Q_0=\frac{P_{n_0}-W/2}{1-W},\qquad
Q_{j+1}=\frac12+\frac{P_{2n_j}-P_{n_j}}{w_j}
\]
have Bernstein coefficients in $[0,1]$, and telescoping gives
\[
(1-W)Q_0+\sum_{j\ge0}w_jQ_{j+1}=f.
\]
Sample this mixture. To simulate the selected polynomial of
degree $m$, flip the input coin $m$ times and let $h$ be the
number of heads. Return a known-bias coin whose bias is the
polynomial's $h$-th Bernstein coefficient. Its unconditional
bias is the selected polynomial evaluated at $p$, so the
mixture is an exact $f(p)$-coin.

The expected number of input flips is at most
\[
n_0+\sum_{j\ge0}2n_jw_j<\infty.
\]
Only one Bernstein coefficient is needed on any invocation.
For a degree-$2n$ branch, compute the single hypergeometric
expectation above using at most $n+1$ terms. Binomial-coefficient
recurrences generate its probabilities in $O(n)$ operations,
and each term requires a constant-work evaluation of $b_n$.
Thus the additional work on that branch is $O(n)$ and has
constant expectation as well. Mixture selection has constant
expected work because the non-base weights are proportional
to $4^{-j}$. All constants depend only on $f$ and $\gamma$,
and the construction is uniform over $p\in[0,1]$.
\end{proof}

\section{Finite precision}\label{sec:finite-precision}

Finite precision suffices for both the deterministic optimizer and
the randomized construction. The optimizer tolerates small errors
in its updates; for the sampler, we instead implement its exact
coin decisions by successively refining numerical bounds.
The work cap then gives a finite computation on every invocation.

We count numerical bit operations, while treating graph access,
vertex identifiers, and the known graph size $n$ in the usual
oracle or word model. Charging also for these $O(\log n)$-bit
words introduces an additional polynomial factor in $\log n$.
Independent random bits and uniform vertex samples are supplied by
an oracle; we do not charge for storing or addressing random seeds.
We may replace the requested accuracy by a dyadic number within a
factor of two, so its description has $O(\log(2/\eps))$ bits.

\subsection{The optimizer and bounded messages}
\label{sec:finite-optimizer}

\begin{proposition}\label{prop:finite-optimizer}
The deterministic construction can be implemented with finite
arithmetic, retaining its $(1-\eps)$ fractional-matching guarantee,
its radius $O(\eps^{-1}\log(2\Delta))$, and its bound
$(2\Delta)^{O(\eps^{-1}\log(2\Delta))}$ on graph queries and
numerical bit operations per edge query.
There is also a distributed implementation using
$T=O(\eps^{-1}\log(2\Delta))$ rounds and
$O(\log(2\Delta/\eps))$ bits per edge per round.
The total computational work at each vertex $v$ is at most
\[
(\deg(v)+1)\operatorname{poly}\!\left(
\frac{\log(2\Delta)}{\eps}\right).
\]
In particular, this is an $O(\eps^{-1}\log(2\Delta))$-round
CONGEST algorithm when $\eps\ge n^{-O(1)}$ and $\Delta\le n$.
\end{proposition}

The useful error bound is a perturbation of the potential proof,
not a comparison between the rounded and exact trajectories.
Write $n_0$ for the number of nonisolated vertices and fix a
numerical tolerance $0<\omega<1$.
Store the transformed price $\tau$ as a rational number, with
the mathematical state price defined by $x=\tanh\tau$.
Thus the state remains interior even when its price is very close
to an endpoint. Numerical evaluations of $x$ need not be exact.
All rounding and evaluation rules are fixed and deterministic.

\paragraph{Errors in the update inequalities.}
Use the notation $a(h)$ and $\MD$ from \Cref{sec:oracles} for
the exact score and mirror-descent map at the current state.
Require each implemented distribution update to have the form
\begin{equation}\label{eq:finite-md}
\widetilde p^+=\MD_p^\beta(a(p)+e),
\qquad \|e\|_\infty\le\omega.
\end{equation}
For the second distribution update of the extragradient stage,
the base is $\bar q$ and the score is $a(q^+)$, as in
\Cref{lem:egoracle}.
Require each implemented price response $\widetilde y=\tanh
\widetilde\tau$ to satisfy
\begin{equation}\label{eq:finite-box}
\bigl\|v^x+2(\mathbf1+Lh)\widetilde\tau
           -2(\mathbf1+Lq)\tau\bigr\|_\infty\le\omega.
\end{equation}
Here $v$ is the gradient vector prescribed by the algorithm at
its current, possibly rounded, state. Errors in evaluating this
vector are included in these residuals.

These conditions add at most $6\omega n_0$ to the right-hand side
of \Cref{lem:gradoracle}, and at most $8\omega n_0$ to that of
\Cref{lem:egoracle}. Indeed, a score error contributes at most
$2\omega n_0$ in each use of the mirror-descent identity, since
two distributions have $\ell_1$ distance at most two per block.
Also,
\[
|\tanh'(t)|\le1,
\qquad
\left|\frac{d}{dt}\psi(\tanh t)\right|
=|t|\operatorname{sech}^2t\le1.
\]
By \eqref{eq:finite-box}, the transformed output differs from
the exact box response by at most $\omega/2$ per coordinate.
Consequently, replacing the exact response by this output
contributes at most $2\omega n_0$ through the proposal gradient
and $2\omega n_0$ through price stationarity.
There is one mirror-descent error in the gradient stage and two
in the extragradient stage, giving the asserted constants.

The proof of \Cref{lem:regret} therefore gives
\begin{equation}\label{eq:finite-regret}
\eta\inner{g(z'_t)}{z'_t-w}
\le2\Phi_t(w)-2\Phi_{t+1}(w)+22\omega n_0.
\end{equation}
Keep the initial proposal distributions exact, and initialize
$\tau_0$ within $\omega$ of $\atanh(-1+1/\Delta)$.
For the comparison points in \Cref{lem:phi0}, this changes
$\Phi_0$ by at most $7\omega n_0$.
To see this, differentiate its price-dependent part with respect
to one initial transformed price: the derivative is
\[
-2(1+\ell_v(q_0))(w^x_v-\tanh\tau_v)
+2(\ell_v(q_0)-\ell_v(q^*))\tau_v\operatorname{sech}^2\tau_v,
\]
whose absolute value is at most seven because
$\ell_v(q_0)\le1/4$ and $\ell_v(q^*)\in\{0,1\}$.
Summing \eqref{eq:finite-regret} exactly as in
\Cref{sec:rate-proof} now yields
\begin{equation}\label{eq:finite-rate}
2\nu_f(G)-P(\widehat q)
\le \frac{3B_\Delta\nu_f(G)+21\omega n_0}{T}
      +33\omega n_0.
\end{equation}
Thus $\omega=c\eps/\Delta$ for a sufficiently small absolute
$c>0$, and $T=C\eps^{-1}\log(2\Delta)$ for a sufficiently
large $C$, leave a constant fraction of the error allowance
for the final output rounding. We use
$n_0\le2\Delta\nu_f(G)$ from \Cref{lem:active}.

\paragraph{Implementing the residual bounds.}
Only modest absolute accuracy is needed in messages, even
though locally stored probabilities may be very small.
Let $\Pi_t=(\mathbf1+Lq_t)\tau_t$.
The main price update and \eqref{eq:finite-box} give
\[
\Pi_{t+1}=\Pi_t+\frac{Lq'_t-\mathbf1}{12}
                 +\frac{r_t}{2},\qquad
\|r_t\|_\infty\le\omega.
\]
Since every load is at most $\Delta$, all transformed prices,
including provisional responses, have magnitude
$O(\log(2\Delta)+T\Delta)$.
The logarithmic-ratio recurrence in \Cref{lem:ratios} still
contracts by $1/4$, with an additional $O(\omega)$ perturbation.
It follows that all update ratios remain between fixed positive
constants and every proposal probability is at least
$e^{-O(T)}/(2\Delta^2)$.

Evaluate each unnormalized proposal weight to relative error
$O(\omega/\beta)$, round it to a positive dyadic number, and
normalize the resulting weights exactly.
If a weight is changed by a factor $e^d$, this is precisely a
score perturbation $-\beta d$; allocating the error allowance
between score evaluation and weight rounding gives
\eqref{eq:finite-md}.
The normalized rational probabilities have numerators and
denominators of $O(T+\log(\Delta/\omega))$ bits.
Elementary functions on these inputs can be evaluated by
convergent series and range reduction in polynomial bit work.
In particular, $\psi(\tanh\tau)$ can be evaluated stably as
$\tau\tanh\tau-\log\cosh\tau$, using
$\log\cosh\tau=|\tau|+\log(1+e^{-2|\tau|})-\log2$.

For communication, round each required scalar $q_{uv}$,
$x_v$, or $\psi(x_v)$ to absolute error at most $h=2^{-b}$,
where
\begin{equation}\label{eq:finite-mesh}
h\le \frac{\omega}{C_1(T+1)(\Delta+1)^4}
\end{equation}
for a sufficiently large absolute $C_1$.
Incoming load errors are at most $\Delta h$.
The price formula divides by $1+\ell\ge1$ and its numerator
uses a transformed price of magnitude $O(T\Delta+\log\Delta)$.
Consequently these message errors, and local rounding on the
same mesh, contribute $O((T+1)(\Delta+1)^3h)$ to the
residual in \eqref{eq:finite-box}.
The preceding Lipschitz bounds give $O(\omega)$ errors in the
scores; choosing smaller constant fractions of the tolerances
gives \eqref{eq:finite-md} as well.
Each round has only a constant number of neighbor exchanges.
Equation~\eqref{eq:finite-mesh} requires
$b=O(\log(T\Delta/\eps))=O(\log(2\Delta/\eps))$ bits.
All larger representations and normalizations remain local.

Each round processes $O(\deg(v)+1)$ scalar entries at
vertex $v$, with representations of
$O(T+\log(\Delta/\eps))$ bits. Each scalar update requires
polynomial bit work in this length.
Summing over $T$ rounds, and including initialization
and final output rounding, gives the claimed total
computational-work bound per vertex.

\paragraph{A feasible finite output.}
For the output average, round each transmitted tentative
probability downward to the $h$-grid and put the missing mass
on the dummy. The resulting average $q^\flat$ satisfies
$q^\flat_{uv}\le\widehat q_{uv}$ on real coordinates and
\[
P(q^\flat)\ge P(\widehat q)-n_0\Delta h.
\]
Its entries have denominator $T2^b$, so its loads and accepted
directed assignments can be computed exactly with
$O(b+\log(T\Delta))$-bit rational numbers.
Round each accepted directed assignment downward to a common
dyadic grid of spacing at most $c\eps/\Delta$, exchange the
two directed values, and average them as in
\eqref{eq:det-fractional}.
This gives exact endpoint agreement and preserves feasibility.
The final decrease in fractional value is at most
$|E|c\eps/\Delta\le2c\eps\nu_f(G)$.
Together with \eqref{eq:finite-rate}, this proves the value
guarantee in \Cref{prop:finite-optimizer}.
The local polynomial bit work is absorbed by the stated LCA
bound. For arbitrary smaller $\eps$, splitting a scalar message
across CONGEST rounds gives the bound
\[
O\!\left(\frac{\log(2\Delta)}{\eps}
\left\lceil\frac{\log(2\Delta/\eps)}{\log n}\right\rceil\right).
\]

\subsection{Exact sampling with finite arithmetic}
\label{sec:finite-sampling}

For the randomized construction, we keep the original exact
iterates as the target distributions.

\begin{proposition}\label{prop:finite-sampling}
With the randomness oracle described above, the uncapped
sampler has an exact implementation with expected numerical
bit work $(2\Delta)^{O(1/\eps)}$.
Capping this work, rather than the number of exact-real
operations, retains all the guarantees and asymptotic bounds
of \Cref{thm:randomized}. Every capped invocation uses finite
precision and terminates within its deterministic work budget.
\end{proposition}

\paragraph{Known numerical biases.}
A coin with a computable bias $a$ does not require computing
$a$ exactly. Compute certified intervals containing $a$ and
reveal bits of an independent uniform random number until the
two intervals are disjoint. With interval widths $O(2^{-r})$,
the probability of remaining undecided is $O(2^{-r})$,
uniformly in $a$. Thus a bias evaluable to error $2^{-r}$
in $\operatorname{poly}(s+r)$ bit operations has an exact coin
implementation with $\operatorname{poly}(s+1)$ expected bit
work. This observation concerns known numerical coefficients
inside a factory; unknown input biases are still accessed
only through independent input flips.

\paragraph{The smooth factory.}
The proof of \Cref{lem:smooth-factory} selects a polynomial of
degree $\Theta(2^j)$ with probability $O(4^{-j})$.
For a degree-$2n$ branch, its only nontrivial coefficient is a
scaled difference involving a hypergeometric expectation of
$b_n(H/n)$. It is enough to approximate this coefficient to
error $2^{-r}$ in
\begin{equation}\label{eq:finite-coefficient-cost}
n\operatorname{poly}(\log(n+1)+r)
\end{equation}
bit operations. This avoids the unnecessary cost of forming
exact large binomial coefficients.

Here is such an evaluation. For
$H\sim\operatorname{Hypergeom}(2n,j,n)$, start at a mode with
unnormalized weight one. Generate the other weights outward
using adjacent probability ratios, which are at most one in
this direction; for example,
\[
\frac{\Prob\{H=k+1\}}{\Prob\{H=k\}}
=\frac{(j-k)(n-k)}{(k+1)(n-j+k+1)}.
\]
There are $O(n)$ terms, and the sum of the unnormalized
weights is at least one. With $s$ fractional bits, rounding
these products and sums introduces $O(n^2 2^{-s})$ error.
The mixture coefficient scales the difference by $O(n^2)$,
so $s=r+O(\log(n+1))$ suffices.
Evaluating $f$ and $f''$ at this precision therefore gives
\eqref{eq:finite-coefficient-cost} whenever their evaluation
cost is polynomial in the argument length and precision.
The known-bias comparison above makes the expected work of
this branch $n\operatorname{poly}(\log(n+1))$.
Summing over branches gives a constant, since
$\sum_{j\ge0}4^{-j}2^j\operatorname{poly}(j+1)<\infty$.

The conversion functions used in \Cref{sec:sampler-exact}
satisfy this evaluation condition. They are fixed affine
rescalings of
\[
F(\xi)=\tanh\frac{\xi}{\sqrt{1-\xi^2}},
\qquad G(\xi)=\psi(F(\xi)),
\]
extended constantly past $\xi=\pm1$.
For $t=\xi/\sqrt{1-\xi^2}$, their deviations from the endpoint
values, and their second derivatives, are bounded by
$C_2(1+|t|)^7e^{-2|t|}$.
Thus when $|t|\ge C_3(r+1)$ one uses the endpoint values
and zero derivatives to precision $2^{-r}$.
Otherwise all elementary evaluations have arguments of
polynomial size in $r$ and the input length.

\paragraph{Elementary formulas.}
For a finite-arithmetic implementation of \Cref{lem:factory},
one may use the preceding smooth factory in place of its
individual nonlinear constructions.
Retain its bounded unbiased representations:
$\E X=a$ and $|X|\le B$ for a fixed known bound $B$.
A draw of $X$, followed by a coin of bias $(X+B)/(2B)$,
supplies a coin of bias $(a+B)/(2B)$.
For each reciprocal, root, or exponential operation, extend
the desired scalar function smoothly outside its promised
interval and encode its output inside a fixed subinterval of
$(0,1)$. Such an extension can be chosen using fixed
polynomial transition functions, with $f$ and $f''$ evaluable
in polynomial bit work. The smooth factory gives its encoded
output coin, which decodes to another bounded unbiased
representation. Addition and multiplication of independent
representations work as before.
Finally, use a smooth extension of the affine decoding map
that agrees with the requested output bias on
$[\gamma,1-\gamma]$.
The formula and all interval bounds are fixed, so this
implementation has constant expected input calls and
additional numerical bit work. It changes no output law.
This assertion concerns the computable constants and fixed
functions in our sampler, not arbitrary real constants in a
general factory specification.

\paragraph{Composition and the work cap.}
The initialization coins have expected
$\operatorname{poly}(\log(2\Delta))$ numerical bit cost.
Applying \Cref{lem:compose} through the same constant-depth
dependencies as in \Cref{prop:cost} bounds a round-$t$
invocation by $A^{t+1}\operatorname{poly}(t+\log(2\Delta))$
expected bit work, for an absolute $A$.
Consequently the exact sampler still has expected work
$(2\Delta)^{O(1/\eps)}$.
Use this bound in \eqref{eq:work-cap}, with the same abort
probability $\tau=\Theta(\eps/\Delta)$, counting all numerical
work and recursive calls against the budget.
The coupling proof of \Cref{lem:cap} is unchanged: an unfinished
proposal becomes the dummy, so real-coordinate probabilities
only decrease. Fixed random streams reproduce the same capped
outputs on every reevaluation.

The marking probability in \eqref{eq:halving-parameters}
is a power of two, so each rounding mark uses exactly
$s=O(\log(2/\eps))$ random bits. This deterministic cost
is included in the rounding bound of \Cref{prop:halving}.
The estimator uses fresh uniform vertex samples and the same
capped proposals. Its integer counts can be accumulated by
binary counters, with amortized constant bit work per increment;
there are at most $\Delta K$ increments per trial.
Hence both the LCA and estimator retain
their stated worst-case numerical work bounds.
The uncapped exact sampler can require arbitrarily fine
precision on rare executions; no fixed precision bound for
that uncapped procedure is claimed or needed.

\end{document}